\documentclass[pdflatex,sn-mathphys-num]{sn-jnl}
\usepackage{graphicx}%
\usepackage{multirow}%
\usepackage{subcaption}
\usepackage{amsmath,amssymb,amsfonts}%
\usepackage{amsthm}%
\usepackage{mathrsfs}%
\usepackage[title]{appendix}%
\usepackage{xcolor}%
\usepackage{textcomp}%
\usepackage{manyfoot}%
\usepackage{booktabs}%
\usepackage{algorithm}%
\usepackage{algorithmicx}%
\usepackage{algpseudocode}%
\usepackage{listings}%
\usepackage{xspace}
\usepackage{tabularx}
\usepackage[many]{tcolorbox}

\theoremstyle{thmstyleone}%
\theoremstyle{thmstyletwo}%

\theoremstyle{thmstylethree}%

\begin{document}

\newcommand{\method}{{\textsc{Cupid}}\xspace}
\newcommand{\rep}{{\textsc{REP}}\xspace}
\newcommand{\siamese}{{\textsc{Siamese Pair}}\xspace}
\newcommand{\sabd}{{\textsc{SABD}}\xspace}
\newcommand{\rr}{{\textit{RR}}\xspace}
\newcommand{\llama}{{\textsc{Llama 3}}\xspace}
\newcommand{\openchat}{{\textsc{OpenChat}}\xspace}
\newcommand{\phithree}{{\textsc{Phi-3}}\xspace}
\newcommand{\xmarkg}{\textcolor{lightgray}{\ding{55}}\xspace}

\title{\method: Leveraging ChatGPT for More Accurate Duplicate Bug Report Detection}


\author{\fnm{Ting} \sur{Zhang}}\email{tingzhang.2019@phdcs.smu.edu.sg}

\author{\fnm{Ivana Clairine} \sur{Irsan}}\email{ivanairsan@smu.edu.sg}

\author{\fnm{Ferdian} \sur{Thung}}\email{ferdianthung@smu.edu.sg}

\author{\fnm{David} \sur{Lo}}\email{davidlo@smu.edu.sg}

\affil{\orgdiv{School of Computing and Information Systems}, \orgname{Singapore Management University}, \orgaddress{\country{Singapore}}}


\abstract{Duplicate bug report detection (DBRD) is a long-standing challenge in both academia and industry.
Over the past decades, researchers have proposed various approaches to detect duplicate bug reports more accurately.
With the recent advancement of deep learning, researchers have also proposed several deep learning-based approaches to address the DBRD task.
In the bug repositories with many bug reports, deep learning-based approaches have shown promising performance.
However, in the bug repositories with a smaller number of bug reports, i.e., around 10k, the existing deep learning approaches show worse performance than the traditional approaches.
Traditional approaches have limitations, too, e.g., they are usually based on the bag-of-words model, which cannot capture the semantics of bug reports.
To address these aforementioned challenges, we seek to leverage a state-of-the-art large language model (LLM) to improve the performance of the traditional DBRD approach.

In this paper, we propose an approach called \method, which combines the best-performing traditional DBRD approach (i.e., \rep) with the state-of-the-art LLM (i.e., ChatGPT).
Specifically, we first leverage ChatGPT under the zero-shot setting to identify essential information on bug reports.
We then use the essential information as the input of \rep to detect duplicate bug reports.
We conducted an evaluation by comparing \method with three existing approaches on three datasets.
The experimental results show that \method achieves state-of-the-art results, reaching Recall Rate@10 scores ranging from $0.602$ to $0.654$ across all the datasets analyzed.
In particular, \method improves over the prior state-of-the-art approach by $5\%$ - $8\%$ in terms of Recall Rate@10 in the datasets.
\method also surpassed the state-of-the-art deep learning-based DBRD approach by up to $82\%$.
Our work highlights the potential of combining LLMs to improve the performance of software engineering tasks.}


\maketitle

\section{Introduction}
\label{sec:introduction}
As software systems become larger and more complex, they inevitably contain bugs.
\textit{Bug reports} are the main channel for users to report bugs to developers.
Most software projects use issue tracking systems, such as Bugzilla~\cite{Bugzilla40:online}, Jira~\cite{JiraIssu14:online} and GitHub~\cite{GitHub24:online}, to manage bug reports and track the progress of bug fixing.
When users find a bug, they can submit a bug report to the issue tracking system.
The developers will then review the bug report and decide whether it is a valid bug report.
If the bug report is valid, the developers will work on fixing the bug.
However, many bug reports are duplicates of the existing bug reports.
For example, in the dataset constructed by Lazar et al.~\cite{lazar2014generating}, duplicate bug reports represent 12.67\% - 23\% out of the total bug reports in a system.
It is crucial to identify duplicate bug reports as soon as possible to avoid wasting developers' time and effort on fixing the same bug multiple times.
To improve the efficiency of bug report management, it is desirable to have an automatic approach to identify duplicate bug reports.

Over the past decades, various duplicate bug report detection (DBRD) approaches have been proposed~\cite{sun2011towards,deshmukh2017towards}.
With the rapid development of deep learning, many deep learning-based approaches have been proposed in recent years~\cite{rodrigues2020soft,he2020duplicate,xiao2020hindbr}.
They have demonstrated superior performance when the bug repositories are large enough to train the deep learning models.
For instance, \sabd~\cite{rodrigues2020soft} achieved over 0.6 in terms of Recall Rate@20 in all the experimented datasets.
One of the common characteristics of these datasets is that they all contain more than $80k$ bug reports and over $10k$ \textit{duplicate} bug reports in the training data, which is large enough to train a deep learning model.
It is well acknowledged that deep learning models require a large amount of data to achieve high precision~\cite{riazi2019deep}.
However, bug repositories in many projects are not large enough to train a deep-learning model.

Based on the dataset provided by Joshi et al.~\cite{joshi2019rapidrelease}, it was discovered that out of the 994 studied GitHub projects that have more than 50 stars and forks, the average number of issues was 2,365.
Additionally, it is interesting to note that many active projects, including those with more than $100k$ stars, have fewer than $10k$ issues. 
For example, till 5th May 2024, both \texttt{ohmyzsh/ohmyzsh}~\cite{ohmyzsho48:online} and \texttt{axios/axios}~\cite{axiosaxi17:online} have less than $5k$ issues each, while \texttt{vuejs/vue}~\cite{vuejsvue14:online} has around $10k$ issues. 
Therefore, we argue that most projects do not have tens of thousands of issues.
The repositories with tens of thousands of issues are considered as \textit{atypical}, while a \textit{typical} repository contains less than or around $10k$ issues.
It is essential to highlight that young and fast-growing projects, although currently having a few issues, require more attention in handling the DBRD challenge. 
For instance, \texttt{Significant-Gravitas/Auto-GPT}~\cite{Signific69:online}, which was initially released on March 30, 2023, now contains less than $3k$ issues while it gets $162k$ stars. 
A recent benchmarking study on DBRD by Zhang et al.~\cite{zhang2022duplicate} also confirms that the performance of deep learning-based approaches loses to information retrieval-based approaches when the bug repositories only contain less or around $10k$ duplicate bug reports.
How to improve the performance of DBRD in the \textit{typical} bug repositories remains an open problem.

Before the development of deep learning, many non-deep learning-based approaches have been proposed~\cite{runeson2007detection,jalbert2008automated,sun2011towards,sun2010discriminative} (we refer to them as ``\textit{traditional} approaches'' in this paper).
Compared to deep learning-based approaches, these approaches are more promising for detecting duplicate bug reports in typical bug repositories.
However, traditional approaches rely on either the vector space model~\cite{runeson2007detection} or the bag-of-words model~\cite{jalbert2008automated}.
These models cannot capture the semantics of bug reports.
We seek to improve the performance of non-deep learning-based approaches by considering the semantics of bug reports.

Recently, large language models (LLMs), e.g., \llama~\cite{llama3modelcard}, Phi-3~\cite{phi-blog,abdin2024phi}, and \openchat~\cite{wang2023openchat}, have achieved outstanding performance in a multitude of natural language processing (NLP) tasks~\cite{devlin2019bert,brown2020language,radford2019language}.
However, leveraging the potential of LLMs to improve DBRD's performance is not trivial.
The most straightforward way is to directly query LLMs on whether two bug reports are duplicates.
However, this is impractical due to the following reasons.

\vspace{0.2cm}\noindent{{\textit{(1) Time-consuming and costly.}}}
To obtain the potential master bug reports to which a given bug report may be duplicated, we must pair it with all the bug reports available in the repository.
When a new bug report is submitted, all previously submitted bug reports are considered duplicate candidates.
It is infeasible to query LLMs to compare the given bug report with all the bug reports in the repository, as the LLMs' response is not instantaneous.
While speeding it up is possible (e.g., by running many queries at once), it quickly gets very costly for LLMs such as ChatGPT~\cite{chatgpt:online}, which operates on a pay-per-use basis for their API usage.

\vspace{0.2cm}\noindent{{\textit{(2) Ignorance of other bug reports in the repository.}}} If a method only compares two bug reports at a time, it will not take into account the information present in the other bug reports stored in the repository.
Therefore, it would be hard to decide the relative order of all the duplicate candidates to recommend the top-$k$ duplicate candidates.
Although one possibility is in addition to querying ChatGPT on whether the bug report pair is duplicated or not, we ask ChatGPT to provide a measure of how confident it is in its answer, expressed as a similarity score or confidence score.
However, without considering the information from other bug reports, the similarity score will be less reliable.


\vspace{0.2cm}
To address the challenges mentioned above, we propose a novel approach that combines the advantages of both traditional DBRD approaches and LLMs.
We present \method, which stands for leveraging \underline{\textsc{C}}hatGPT for d\underline{\textsc{up}}l\underline{\textsc{i}}cate bug report \underline{\textsc{d}}etection.
\method aims to tackle the abovementioned challenges when directly querying LLMs for DBRD.
We propose to leverage LLMs as an intermediate step to improve the performance of the traditional DBRD approach.
Based on the recent benchmarking study by Zhang et al.~\cite{zhang2022duplicate}, \rep~\cite{sun2011towards} demonstrates the best performance in the datasets with a typical number of issues, which is also the focus of this work.
Thus, we select \rep as the backbone duplicate retrieval method.
Specifically, \method leverages state-of-the-art ChatGPT to extract keywords from bug reports and then incorporate them with \rep to achieve better performance.
By doing so, \method avoids using ChatGPT to compare the given bug report with all the bug reports in the repository.
Furthermore, by standing on the shoulder of the traditional DBRD approach, \method also takes the information of the other bug reports in the repository into consideration.
In particular, $BM25F_{ext}$ used by \rep calculates inverse document frequency (IDF), which is a global term-weighting scheme across all the bug reports.
Our contribution can be summarized as follows:
\begin{itemize}
    \item{\textbf{Approach:} We propose \method, which combines modern LLMs with the traditional DBRD technique to enhance the accuracy of DBRD in software systems with the typical number of bug reports.}
    \item{\textbf{Evaluation:} We evaluate \method on three datasets from open-source projects and compare \method with three prior state-of-the-art approaches. The experimental results indicate that \method surpasses the performance of these existing DBRD approaches. Notably, \method achieves RR@10 scores ranging from $0.602$ to $0.654$ across all the datasets analyzed.}
    \item{\textbf{Direction:} We show that leveraging ChatGPT indirectly in conjunction with existing approaches can be beneficial. We anticipate that this will pave the way for future research to explore innovative ways to utilize state-of-the-art techniques with traditional ones.}
\end{itemize}

The structure of this paper is as follows.
Section~\ref{sec:background} introduces the background of DBRD.
Section~\ref{sec:approach} presents the details of \method.
We describe the experimental design in Section~\ref{sec:experimental_design}.
Section~\ref{sec:results} presents the experimental results.
We discuss the threats to validity in Section~\ref{sec:discussion}.
Finally, Section~\ref{sec:conclusion} concludes this paper and discusses future work.

\section{Background}
\label{sec:background}

In this section, we first introduce the essential concepts, including bug reports and duplicate bug reports, and finally, we discuss the task of DBRD.

\textit{Bug reports} are the primary means for users to communicate a problem or request features to developers~\cite{bettenburg2008makes}.
Software projects usually rely on \textit{issue tracking systems} to collect these bug reports.
While the supported fields can vary from system to system, the \textit{textual} information is included in all the issue tracking systems.
The \textit{textual} fields in a bug report usually consist of a summary (title) and a description.
In Bugzilla or Jira, there are also several \textit{categorical} fields, such as \texttt{priority} (bug assignees use this to prioritize their bug), \texttt{product}, \texttt{component}, etc.

DBRD aims to correctly link a duplicate bug report towards its \textit{master} bug report. 
Following prior works~\cite{runeson2007detection,sun2011towards,rodrigues2020soft}, we denote the first submitted bug report on a specific fault as the \textit{master} bug report and the subsequent bug reports on the same fault as \textit{duplicates}.
All the bug reports which are duplicates of each other, including \textit{master} and \textit{duplicates}, are in the same \textit{bucket}.
To help understand these concepts easier, we can imagine a \textit{bucket} as a Hash Table, where the key is the master bug report, while the values are duplicate bug reports and themselves.
Thus, for a unique bug, both the key and value would be just itself.

In the literature, depending on the problem setting, DBRD has been evaluated in two manners, i.e., (1) \textit{classification} and (2) \textit{ranking}.
In the \textit{classification} manner, the task is to classify whether two bug reports are duplicates or not.
\textsc{DC-CNN}~\cite{he2020duplicate} and \textsc{HINDBR}~\cite{xiao2020hindbr} are two recent endeavors in this manner.
In the \textit{ranking} manner, the task is to rank the candidate bug reports according to their similarity to the given bug report.
Referring to the Hash Table metaphor earlier, given a newly submitted bug report, DBRD technique finds the bucket to which it belongs (also equivalent to linking duplicate bug reports to its master).
If it does not belong to any existing bucket, a new bucket in which the key and value are itself should be created. 
In this work, we adopt the \textit{ranking} manner, which is more practical in real-world applications~\cite{rodrigues2020soft,zhang2022duplicate}.

In the past decades, researchers have proposed various approaches to address the DBRD task in the ranking manner~\cite{wang2008approach,nguyen2012duplicate}.
Different DBRD approaches mainly differ in 
(1) feature engineering: which features in bug reports are selected and how these features are represented, and (2) similarity measurement: how to measure the similarity between two bug reports~\cite{sureka2010detecting}.
In terms of \textit{feature engineering}, we further break it down into two parts: (1) what features are selected and (2) how to represent the features.
All existing methods use textual information and most of them use categorical information.
Textual features, i.e., summary and description, include the most useful information about a bug.
Different methods differ in which categorical features to use.
To model these features, traditional methods utilize bag-of-words, character-level N-gram, or $BM25_{ext}$ to model textual features~\cite{jalbert2008automated,sun2011towards,sureka2010detecting}, while bag-of-words or hand-crafted methods are usually used to model categorical features.
Deep learning-based methods utilize word embeddings, such as GloVe~\cite{pennington2014glove} and word2vec~\cite{mikolov2013efficient}, to represent the textual information. Other types of neural networks, such as HIN2vec~\cite{fu2017hin2vec}, are used to represent categorical information.
For \textit{similarity measurement}, traditional methods usually use Cosine, Dice, and Jaccard similarity~\cite{jalbert2008automated,runeson2007detection}.
While some deep learning-based models also adopt this similarity measure~\cite{deshmukh2017towards}, some of them leverage neural networks to learn the similarity~\cite{rodrigues2020soft}.

\section{Approach}
\label{sec:approach}

\begin{figure*}[t]
	\includegraphics[width=1\linewidth]{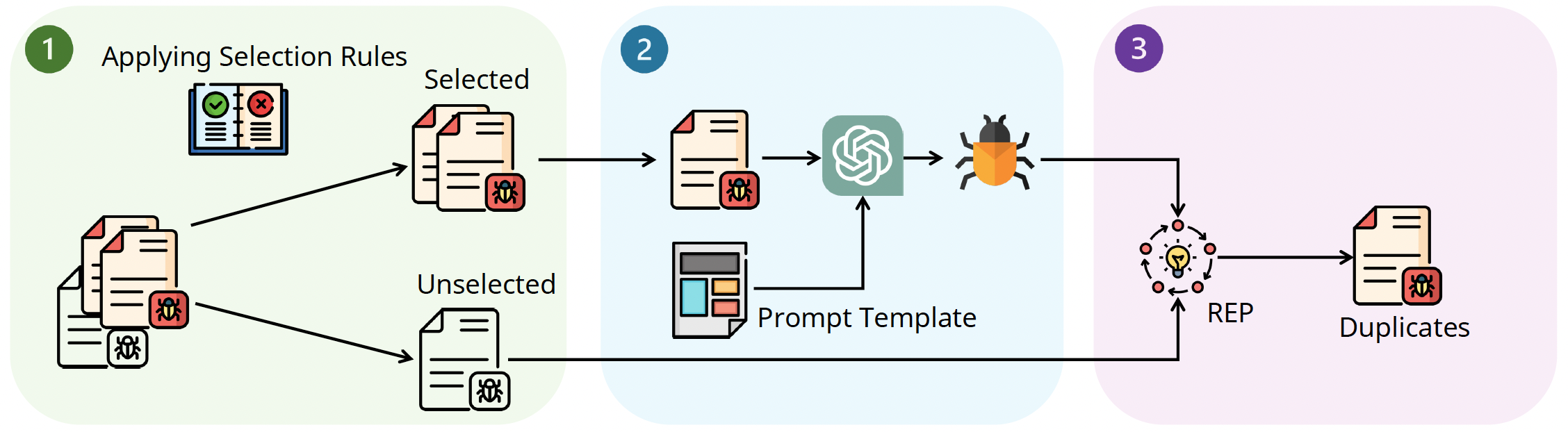}
	\caption{\method contains three stages: In Stage 1, it applies selection rules to select the test bug reports that need to be processed; In Stage 2, it utilizes ChatGPT to process the selected bug reports; In Stage 3, it leverages \rep to retrieve potential master bug report for each test bug report.}
	\label{fig:method_overview}
\end{figure*}

We propose \method to combine the advantages of both the traditional DBRD approach and LLMs.
As mentioned earlier, our work focuses on solving the DBRD challenge in the repositories with a typical number of issues; evidence shows that traditional DBRD approaches would fit more in this condition than deep learning-based approaches~\cite{zhang2022duplicate}.
Figure~\ref{fig:method_overview} shows an overview of the proposed method.
The overall process consists of three main stages: (1) Applying \textit{selection rules} to select the bug reports that need to be processed by ChatGPT,
(2) Running ChatGPT with \textit{prompt template} to get the essential keywords of the selected bug reports, 
and (3) Applying \rep to retrieve \textit{potential master bug reports}.

\subsection{Applying Selection Rules}
\begin{table}[t]
    \centering
    \caption{Example content can be matched by the dotted identifier regular expression.}
    \begin{tabular}{@{}p{\columnwidth}@{}}
    \toprule
    Code snippet: \\
    \texttt{SELECT * FROM \textbf{database.schema.table;}} \\
    \midrule
    Error message: \\
    \texttt{[ERROR] Failed to execute goal} \\
    \texttt{\textbf{org.apache.maven.plugins:maven-javadoc-plugin:}} \\
    \texttt{\textbf{3.0.1:javadoc} (default-cli) on project hadoop-hdfs: An error has occurred in Javadoc report generation:} \textit{... omitted}\\
    \midrule
    Stack traces: \\
    \texttt{\textbf{java.lang.NullPointerException} at} \\
    \texttt{com.example.MyClass.myMethod(MyClass.java:42)} \textit{... omitted} \\
    \bottomrule
    \end{tabular}
    \label{tab:selection_rule}
\end{table}

Considering the computational cost of ChatGPT, we did not run ChatGPT on all the bug reports in the test dataset.
Similarly, in practice, we do not need to run ChatGPT on each newly submitted bug report.
To further improve efficiency while keeping accuracy, we explore and propose selection rules.
These rules are based on the content of the bug reports to prioritize bug reports that are harder to process by \rep while reducing the number of bug reports that are fed into ChatGPT. 
The selection criterion is based on the content. 
We select the bug reports whose descriptions contain dotted identifiers or URLs.
We use regular expressions \verb|r"\w+\.\w+\.\w{1,}"| and \verb|r"https?://\S+"| to match and select these bug reports.
The dotted identifiers usually exist in fully qualified class names, namespaced identifiers, error messages, and stack traces.
Table~\ref{tab:selection_rule} shows examples of the content that can be matched by the dotted identifier regular expression.
For developers, this information is useful.
However, for a DBRD method, this information can be hard to process.
We select these bug reports because not all the code snippets are useful for \rep to retrieve the potential master bug reports. 
We also do not directly remove code snippets. 
The reason is that we want to keep the original structure of the bug reports so ChatGPT can understand the language better.
We then utilize ChatGPT to identify keywords from these bug reports.

\subsection{Running ChatGPT with Prompt Template}
After Stage 1, we run ChatGPT on the selected bug reports, i.e., the description contains dotted identifiers or URLs.
Prompt~\cite{liu2023pre} is a set of instructions that can be used to probe LLMs to generate the target outcome~\cite{bang2023multitask}.
Prior studies have empirically shown that LLMs are sensitive to prompts~\cite{zhang2023revisiting,zhang2023sentiment}.
Thus, for different tasks, the prompts should be carefully designed to enable LLMs to demonstrate their abilities.
To get the most suitable prompt for ChatGPT, we directly query ChatGPT to get the prompt.
Table~\ref{tab:prompt_1} shows the query we used to get the initial prompt and the prompt replied by ChatGPT.
The latter one is what we used in \method.
After getting the response from ChatGPT, we replace the original summary and description in the bug report with the returned identified keywords for summary and description.
We keep the remaining part of the bug report unchanged.

\begin{table}[t]
\centering
\caption{Query to ChatGPT to Generate Initial Prompt Template and ChatGPT's Response.}
\begin{tabular}{@{}p{1\columnwidth}@{}}
\toprule
\textsf{\textbf{To Get Initial Prompt Template}}\\
    Provide a concise prompt or template that can be used to identify keywords from the summary and description of a bug report. These keywords will be used as input to detect duplicate bug reports. The output format will be:
    \\
    Summary: [Selected Keywords]
    \\
    Description: [Selected Keywords]
    \\
\midrule
\textsf{\textbf{Prompt Template}}\\
Identify keywords from the summary and description of the bug report that can be used to detect duplicates.\\
Output format: 
\\
Summary: [Selected Keywords]
\\
Description: [Selected Keywords]
\\\\
Summary: \{\{Summary\}\}
\\
Description: \{\{Description\}\}
\\
\bottomrule
\end{tabular}
\label{tab:prompt_1}
\end{table}

\subsection{Retrieving Potential Master Bug Reports}
\label{sec:rep-description}
Considering the superiority of \rep in the task of DBRD shown in a recent study~\cite{zhang2022duplicate}, especially on projects with a typical number of issues, we use \rep as the DBRD approach in \method.
Here, we briefly introduce the \rep approach to make the paper self-contained.
We refer the readers to the original paper~\cite{sun2011towards} for more details.

As shown in Formula~\ref{eq:rep}, \rep is a linear combination of seven features, including \textbf{textual} features and \textbf{categorical} features.

\begin{align}
\label{eq:rep}
\begin{aligned}
    \text{REP}(d, q) = \sum_{i=1}^{7} w_i \cdot \text{feature}_i
\end{aligned}
\end{align}

\noindent, where $d$ is the bug report in the repository $R$, $q$, is the query (i.e., new bug report), $w_i$ is the weight of the $i$-th feature, and $feature_i$ is the $i$-th feature.
The first two features are both \textbf{textual} features, and the rest five features are \textbf{categorical} features.
Formula~\ref{fig:features} shows how to get each feature.
\begin{align}
\label{fig:features}
\begin{aligned}
& \text{feature}_1(d, q)=\text{BM25F}_{e x t}(d, q) \text { //of unigrams } \\
& \text{feature}_2(d, q)=\text{BM25F}_{e x t}(d, q) \text { //of bigrams } \\
& \text{feature}_3(d, q)= \begin{cases}1, & \text{if } d \cdot p r o d=q \cdot p r o d \\
0, & \text{otherwise }\end{cases} \\
& \text{feature}_4(d, q)= \begin{cases}1, & \text{if } d . comp=q . comp \\
0, & \text {otherwise }\end{cases} \\
& \text{feature}_5(d, q)= \begin{cases}1, & \text{if } d.type = q.type \\
0, & \text {otherwise }\end{cases} \\
& \text{feature}_6(d, q)=\frac{1}{1+\mid d.prio - q.prio \mid} \\
& \text{feature}_7(d, q)=\frac{1}{1+\mid d.vers - q.vers \mid} \\
\end{aligned} 
\end{align}

The first two features regard the textual similarity between two bug reports over the fields summary and description.
These two textual features are calculated by $BM25F_{ext}$ between bug report $d$ and query bug report $q$.
$BM25F$~\cite{robertson2004simple,zaragoza2004microsoft} is an effective textual similarity function for retrieving documents with structures.
The authors of \rep extend $BM25F$ by considering term frequencies in queries and proposed $BM25F_{ext}$.


In $feature_1$, summary and description are represented in uni-gram, while in $feature_2$, summary and description are represented in bi-gram.
Thus, the input of $BM25F_{ext}$ consists of a bag of uni-grams and bi-grams in both features.
For $feature_{3-5}$, they are the categorical features of \texttt{product}, \texttt{component}, and \texttt{type}, respectively.
If the corresponding field value from $d$ and $q$ is the same, the value of the feature is 1, otherwise, it is 0.
For $feature_{6-7}$, they are the categorical features of \texttt{priority} and \texttt{version}, respectively.
They are calculated by the reciprocal of the distance between the corresponding field value from $d$ and $q$. 
Overall, the \rep approach contains $19$ free parameters with different initial values.
These parameters are tuned by gradient descent.

\section{Experimental Design}
\label{sec:experimental_design}
In this section, we first outline the research questions we aim to investigate. 
Secondly, we describe the datasets utilized in our study. 
Thirdly, we discuss the evaluation metrics employed. 
Fourthly, we provide succinct descriptions of the baseline models. 
Fifthly, we delve into the six alternatives to ChatGPT considered, with half being open-source LLMs and the other half being statistics-based keyword extraction methods. 
Lastly, we mention the implementation details for running ChatGPT and its alternatives.

\subsection{Research Questions}
To understand whether \method performs better compared to existing state-of-the-art approaches and whether each component of \method is useful, we answer the following two research questions (RQs).

\noindent{\textbf{RQ1:} \textit{How effective is \method compared to the state-of-the-art DBRD approaches?}}
We compare \method with state-of-the-art DBRD techniques, which consider DBRD as a ranking problem, i.e., \rep~\cite{sun2011towards}, \siamese~\cite{deshmukh2017towards}, and \sabd~\cite{rodrigues2020soft}.

\vspace{0.2cm}\noindent{\textbf{RQ2:} \textit{How effective are the components of \method?}}
To answer this RQ, we conduct an ablation study on \textsc{Spark} dataset to investigate the effectiveness of the components of \method.
This RQ is further divided into the following sub-RQs:

\begin{itemize}
    \item \textbf{RQ2.1:} \textit{How effective is the selection rule?} 
    Besides content, we also experimented with selecting bug reports whose description is considered to be \textit{long}. 
    We consider bug reports whose description is longer than \textit{n} words as \textit{long} bug reports. 
    We get \textit{n} by calculating the 75th percentile of the description length in the training set. The reason why we select long bug reports is that long bug reports are usually not concise and contain long stack traces and code snippets. 
    These long bug reports would make it challenging for \rep to retrieve the potential master bug reports.
    We investigate the impact of the length and content of the bug report description on the performance of \method.
    \item \textbf{RQ2.2:} \textit{How effective is ChatGPT compared to other LLMs and statistics-based keyword extraction methods?} We select three open-source LLMs, i.e., \llama (i.e., \texttt{meta-llama/Meta-Llama-3-8B-Instruct} in Hugging Face library~\cite{wolf-etal-2020-transformers}), \phithree (i.e., \texttt{microsoft/Phi-3-mini-128k-instruct}), and \openchat (i.e., \texttt{openchat/openchat-3.5-0106}) based on their performance in MMLU benchmark on the LMSys chatbot Arena leaderboard in May 2024~\cite{LMSysCha85:online}.
    All these three LLMs are open-source and have less than 10B parameters. For the statistics-based keyword extraction methods, we select  \textsc{TFIDF}~\cite{rose2010automatic}, \textsc{KP-Miner}~\cite{el2009kp}, and \textsc{YAKE}~\cite{campos2020yake} based on their good performance on a recent survey~\cite{papagiannopoulou2020review} and popularity in the literature. 
    \item \textbf{RQ2.3:} \textit{How effective is the prompt template?} We investigate the impact of the prompt template on the performance of \method. To get alternative prompts for ChatGPT, we queried ChatGPT again. The prompt we used to get alternative prompts and the prompts replied by ChatGPT are shown in Table~\ref{tab:prompt_ablation}.
\end{itemize}

\begin{table}[t]
    \centering
    \caption{Query to ChatGPT to Generate Alternative Prompt Templates and ChatGPT's Responses.}
    \begin{tabular}{@{}p{1\columnwidth}@{}}
    \toprule
    \textsf{\textbf{To Get Alternative Prompt Templates}} \\
    Provide two alternative prompts: one is more concise, and the other is more verbose. The current prompt template is:
    \\\\
    Identify keywords from the summary and description of the bug report that can be used to detect duplicates.
    \\\\
    Output format:
    \\
    Summary: [Selected Keywords]
    \\
    Description: [Selected Keywords]
    \\
    \midrule
    \textsf{\textbf{More Concise Prompt Template}}\\
    Identify keywords from the bug report to detect duplicates.
    \\\\
    Output format:
    \\
    Summary: [Selected Keywords]
    \\
    Description: [Selected Keywords]
    \\\\
    Summary: \{\{Summary\}\}
    \\
    Description: \{\{Summary\}\}
    \\
    \midrule
    \textsf{\textbf{More Verbose Prompt Template}}\\
    Review the summary and description of the bug report to identify specific keywords that can be used as criteria for detecting duplicate reports. Consider the language used, technical terms, and any unique identifiers mentioned in the report.
    \\\\
    Output format:
    \\
    Summary: [Selected Keywords]
    \\
    Description: [Selected Keywords]
    \\\\
    Summary: \{\{Summary\}\}
    \\
    Description: \{\{Description\}\}
    \\
    \bottomrule
    \end{tabular}
    \label{tab:prompt_ablation}
\end{table}

\subsection{Dataset}
As mentioned in Section~\ref{sec:introduction}, we are concerned about boosting the performance of DBRD, especially in the bug repositories with the typical number of issues.
Therefore, the target datasets are those that contain a typical number of issues.
We employ three datasets, i.e., Spark, Hadoop, and Kibana datasets, which are provided by a recent benchmarking study by Zhang et al.~\cite{zhang2022duplicate}.
These datasets contain around 10k issues each, which is considered a typical number of issues.
These datasets are recent issues, ranging from 2018 to 2022, which addressed the \texttt{age} bias, i.e., the model performs differently on the recent data and old data.
Spark and Hadoop are two popular open-source distributed computing frameworks.
They both use Jira as their issue tracking system.
Kibana is a visualization tool for Elasticsearch, and it uses GitHub as its issue tracking system.
The statistics of the datasets are shown in Table~\ref{tab:dataset}.
The duplicate and non-duplicate pairs were sampled by Zhang et al.~\cite{zhang2022duplicate}.
Their ratio is 1:1.
We obtained the data in the dataset provided by Zhang et al.
In our experiment, we fixed the number of training and validation pairs. 
The number of duplicate bug reports in the test set is the bug reports we investigate.
We report the performance of each approach in terms of how they perform in retrieving the master bug reports for the duplicate bug reports in the test set.

\begin{table}[t]
	\centering
	\caption{Dataset statistics.}
	\label{tab:dataset}
	{
    \begin{tabular}{lrrrrrr}
    \toprule
    \multirow{2}{*}{\bf Dataset} & \multirow{2}{*}{\bf \# BRs} &   \multicolumn{2}{c}{{\bf \# Pairs}} & \multicolumn{2}{c}{{\bf Test}}\\
    \cmidrule{3-6}
    & & {\bf Train} & {\bf Valid} & {\bf Total BRs} & {\bf Dup. BRs} \\
    \midrule
    Spark & 9,579 & 625 & 26 & 2,841 & 81 \\
    Hadoop & 14,016 & 626 & 26 & 3,740 &  92 \\
    Kibana &  17,016 & 724 & 28 & 7,167 &  184 \\
    \bottomrule
    \end{tabular}
	}
\end{table}

\subsection{Evaluation Metrics}
Following prior works on DBRD~\cite{zhang2022duplicate,amoui2013search,deshmukh2017towards,runeson2007detection}, we only use Recall Rate@$k$ (RR@$k$) as the evaluation metric, where $k$ represents the number of bug reports to be considered.
Note that a few other works have also adopted Mean Average Precision (MAP) in the DBRD literature.
However, since MAP considers all of the predicted positions, it is not suitable for our case, where only the top $k$ predictions matter.
This is based on real-world practice, where developers are more likely to check the top $k$ predictions rather than all of the predictions.
A survey on practitioners' expectations towards fault localization also shows that around 98\% of respondents are not willing to check the predictions beyond the top-10 to find the faulty element~\cite{kochhar2016practitioners}.
Additionally, the DBRD feature adopted in the microsoft/vscode repository also only shows at most top-5 predictions~\cite{zhang2022duplicate}.

Furthermore, as already discussed in early work~\cite{runeson2007detection}, the two widely used metrics in information retrieval, i.e., Precision@$k$ and Recall@$k$, do not fit into how DBRD works.
For each query bug report, we only have a master bug report to look for (i.e., the relevant item is only 1).
Consider $k=10$, a successful prediction would lead to Precision@10=1/10(10\%), and Recall@10=1/1(100\%); a failed prediction would lead to Precision@10=0 and Recall@10=0.
Therefore, we adopt RR@$k$ as the evaluation metric.

Following the definition in prior works~\cite{schutze2008introduction,sun2011towards,rodrigues2020soft,wang2008approach}, RR@$k$ is defined as the percentage of duplicate bug reports that are correctly assigned to the bucket they belong to when a model makes a top-$k$ prediction for each test bug report.
In our experiment, RR@$k$ will measure how well DBRD techniques correctly link the duplicate bug reports to their master bug report. 
A higher RR@$k$ indicates that more bug reports in the test set are correctly linked to the bucket they belong to when a model retrieves top-$k$ prediction.

\begin{align}
\label{eq:recall}
\text{Recall Rate} = \frac{N_{recalled}}{N_{total}}
\end{align}

Formula~\ref{eq:recall} shows how to calculate the Recall Rate. 
$N_{recalled}$ refers to the number of duplicate bug reports whose bucket (master bug report) are in the suggested list (with a size of [1,2,...,$k$]).
$N_{total}$ refers to the number of duplicate bug reports investigated.
Considering different sizes of the suggested list, i.e., $k$, we can get RR@$k$.
Following the benchmarking work by Zhang et al.~\cite{zhang2022duplicate}, 
in our case, we consider at most 10 predictions, i.e., $k=[1,2...,10]$.


\subsection{DBRD Baselines}
\noindent{{\bf \rep~\cite{sun2011towards}} The details of \rep can referred in Section~\ref{sec:rep-description}.}

\vspace{0.1cm}\noindent{{\bf \siamese~\cite{deshmukh2017towards}} is the first approach that leverages deep learning for DBRD.
As its name suggests, \siamese utilizes Siamese variants of Convolutional Neural Networks (CNN) and Recurrent Neural Networks (RNN) trained on max-margin objective to distinguish similar bugs from non-similar bugs.
\siamese adopts word embedding to represent the textual data as numerical vectors.
Then, it employs three different types of neural networks to encode \texttt{summary}, \texttt{description}, and categorical features according to their properties. 
Specifically, \texttt{summary} is encoded by bi-LSTM, \texttt{description} is encoded by a CNN, while the categorical information is encoded with a single-layer neural network.
When a new bug report comes, \siamese encodes the bug report with the trained model.
It then calculates the cosine similarity between the new bug report and each bug report in the master set and gives top k predictions.
}

\vspace{0.1cm}\noindent{{\bf \sabd~\cite{rodrigues2020soft}} is the latest deep learning-based DBRD approach.
It comprises two sub-network modules, each comparing textual and categorical data from two bug reports.
In the textual sub-network, the soft-attention alignment mechanism~\cite{parikh2016decomposable} compares each word in a bug report with a fixed-length representation of all words in the other bug report.
By doing so, \sabd learns the joint representation of bug reports.
In the categorical sub-network, each categorical field relates to a lookup table that links the field value to a real-valued vector.
The output vector from each sub-module is concatenated and fed to a fully connected layer.
Finally, a classifier layer, which is a logistic regression, produces the final prediction, i.e., whether the two bug reports are duplicates.
}

\subsection{ChatGPT Alternatives}
\subsubsection{Open-source LLMs}
\noindent{\bf \llama~\cite{llama3modelcard}} is the latest LLMs developed and released by Meta.
\llama is an auto-regressive LLM that uses an optimized Transformer architecture. The tuned versions use supervised fine-tuning and reinforcement learning with human feedback to align with human preferences. It comes in two sizes, i.e., 8B and 70B versions, both of which use grouped-query attention for improved inference scalability.
The \llama instruction-tuned models are optimized for dialogue use cases and outperform many of the available open-source chat models on common industry benchmarks.

\vspace{0.1cm}\noindent{{\bf \phithree}~\cite{abdin2024phi}} is a family of open-source LLMs developed by Microsoft. According to Microsoft, the \textsc{Phi-3-mini} model, with ``only'' 3.8 billion parameters, outperforms models twice its size, while \textsc{Phi-3-small} and \textsc{Phi-3-medium} surpass the performance of much larger models. 
\textsc{Phi-3-mini} is available in two context-length variants: 4K and 128K tokens.
Notably, it is the first model in its class to support a context window of up to 128K tokens, with minimal impact on quality. 
\textsc{Phi-3-mini} are instruction-tuned and trained to follow different types of instructions.

\vspace{0.1cm}\noindent{{\bf \openchat}\cite{wang2023openchat}} are trained on general supervised fine-tuning data, consisting of a small amount of expert data mixed with a large proportion of sub-optimal data, without any preference labels. Despite this, \openchat models deliver exceptional performance on par with ChatGPT, even with a 7B model that can be run on consumer-grade GPUs (e.g., RTX 3090). \openchat introduces the C(onditioned)-RLFT technique, which regards different data sources as coarse-grained reward labels and learns a class-conditioned policy to leverage complementary data quality information. Upon its release in January 2024, \texttt{OpenChat-3.5-0106} was the strongest 7B language model in the world, ranking as the top 7B model on the LMSys Arena~\cite{LMSysCha85:online}, and the top open-source language model not originating from a company.

\subsubsection{Statistics-based Keyword Extraction Methods}
\vspace{0.2cm}\noindent{{\bf \textsc{TF-IDF}~\cite{rose2010automatic}}} serves as the common baseline for the keyword extraction task. \textsc{TF-IDF} scores and ranks phrases according to the formula: $\text{TF-IDF} = \text{Tf} \times \text{Idf}$, where $\text{Tf}$ is the raw phrase frequency, and $\text{Idf} = \log_2 \frac{N}{1 + |d \in D : \text{phrase} \in d|}$. In the $\text{Idf}$ formula, $N$ represents the number of documents in the document set $D$, and $|d \in D : \text{phrase} \in d|$ denotes the number of documents in which the phrase appears.

\vspace{0.2cm}\noindent{{\bf \textsc{KP-Miner}~\cite{el2009kp}}} not only utilizes term frequency and inverse document frequency scores, but also incorporates various other statistical measures. It employs an efficient candidate phrase filtering process and uses a scoring function. 
Specifically, the system retains candidate words that are not separated by punctuation or stop words, considering the minimum allowed visible frequency factor and a cutoff constant defined based on the number of words after the first occurrence of a phrase. 
The system then ranks candidate phrases based on their $\text{Tf}$ and Idf scores, as well as the term position and the lift factor of the compound term relative to its individual terms.

\vspace{0.2cm}\noindent{{\bf \textsc{YAKE}~\cite{campos2020yake}}} utilizes novel statistical metrics to capture contextual information and the spread of terms within the document, in addition to considering the position and frequency of terms. First, \textsc{YAKE} preprocesses the text by splitting it into separate terms. 
Second, a set of five features is calculated for each individual term: casing, word position, word frequency, the number of different terms appearing on the left and right sides of the word, and how often a candidate word appears within different sentences. All these features are then utilized to calculate the $S(w)$ score for each term. 
Finally, a 3-gram sliding window generates a continuous sequence of 1-gram, 2-gram, and 3-gram candidate keywords.

\subsection{Implementation}
\noindent{\textbf{ChatGPT.}} Given that ChatGPT is still fast evolving, it has undergone several iterations~\cite{ChatGPT4:online}.
In this study, we worked on the GPT-3.5 Turbo version.
To interact with ChatGPT, we used its official API.
Although GPT-4 is also available, it is much more expensive than GPT-3.5.
Therefore, we chose to use the cheaper version of ChatGPT, which we believe to have a wider range of users compared to GPT-4.
As such, our results would be more valuable as they apply to a wider range of users.

Since ChatGPT may generate different answers for the same query, we set the temperature to 0 and fixed the random seed to 42 to make the answers more stable.
Even so, we still observed some variations in the answers generated by ChatGPT.
Thus, we ran ChatGPT five times for each bug report and ran the \rep five times to get the average results.
Specifically, we average the RR@$k$ of the five runs of \method.

\vspace{0.2cm}\noindent{\textbf{Open-source LLMs.}} We used the Hugging Face Transformers library~\cite{wolf-etal-2020-transformers} to interact with the open-source LLMs.
Since we set \texttt{do\_sample=False}, the answers generated by the open-source LLMs are stable and do not change across runs.
The LLM models are run only once for each bug report.
We set the \texttt{max\_new\_tokens=2,048}, which means the maximum number of tokens to generate is 2,048.
We use the same prompt templates across different LLMs.

\vspace{0.2cm}\noindent{\textbf{Statistics-based keyword extraction methods.}} We used the open-source keyword extraction toolkit~\cite{boudin:2016:COLINGDEMO} to run the methods.
Since \textsc{TF-IDF}, \textsc{KP-Miner}, and \textsc{YAKE} are deterministic, the keywords extracted by these methods are stable and do not change across runs.
Thus, we run them once for each bug report.
As \textsc{TF-IDF} and \textsc{KP-Miner} rely on the document frequency of the terms, we used the training set to calculate the document frequency of the terms.
We leave the hyper-parameters of these keyword extraction methods as default.

\section{Results}
\label{sec:results}

\subsection{RQ1: Comparing with baselines}
\begin{table}[t]
\centering
\caption{Recall Rate@$k$ obtained on the Spark, Hadoop, and Kibana datasets.}
\label{tab:rq1_result}
{
\begin{tabular}{lrrrrrr}
\toprule
& {\bf RR@1} & {\bf RR@2} & {\bf RR@3} & {\bf RR@4} & {\bf RR@5} & {\bf RR@10}\\
\midrule
{\it Spark} \\
\midrule
\rep & 0.346 & 0.383 & 0.457 & 0.481 & 0.481 & 0.556  \\
\siamese &  0.037 & 0.049 & 0.059 & 0.064 & 0.074 & 0.121 \\
\sabd & 0.202 & 0.247 & 0.281 & 0.294 & 0.304 & 0.331 \\
\midrule
\method & {\bf 0.360}	& {\bf 0.435} &	{\bf 0.472} &	{\bf 0.499}	& {\bf 0.519}	& {\bf 0.602} \\
\midrule
{\it Hadoop} \\
\midrule
\rep &  0.402 & 0.489 & {\bf 0.522} & {\bf 0.554} & 0.576 & 0.609 \\
\siamese &  0.033 & 0.046 & 0.057 & 0.063 & 0.076 & 0.093 \\
\sabd & 0.215 & 0.267 & 0.293 & 0.304 & 0.324 & 0.411 \\
\midrule
\method &  {\bf 0.420} &	{\bf 0.496} &	{\bf 0.522} &	0.546 &	{\bf 0.587} &	{\bf 0.637} \\
\midrule
{\it Kibana} \\
\midrule
\rep & 0.364 & 0.440 & 0.527 & 0.560 & 0.587 & 0.620 \\
\siamese &  0.020 & 0.036 & 0.050 & 0.063 & 0.076 & 0.092 \\
\sabd & 0.293 & 0.382 & 0.428 & 0.467 & 0.489 & 0.555 \\
\midrule
\method & {\bf 0.389} &	{\bf 0.490} &	{\bf 0.557}	& {\bf 0.580}	& {\bf 0.605} &	{\bf 0.654} \\
\bottomrule
\end{tabular}
}
\end{table}

Table~\ref{tab:rq1_result} shows the results of \method and the baselines on the Spark, Hadoop, and Kibana datasets.
Overall, \method consistently improves the DBRD performance in terms of RR@10 on all three datasets, yielding an improvement of $8\%$ (Spark), $5\%$ (Hadoop), and $6\%$ (Kibana) over the prior state-of-the-art approach \rep.
This improvement is obtained by successfully utilizing the content generation ability of ChatGPT to transform the bug reports into a format where only essential information is kept. 
Compared with the best-performing deep learning-based approach, i.e., \sabd, we observe an improvement of up to $82\%$ on the Spark dataset.
In the low-volume datasets, \sabd and \siamese lose to non-deep learning approaches, i.e., \rep and \method.

Comparing the performance of \siamese and \sabd in all three datasets, we can find that \siamese suffers more from the challenge of limited training data.
\siamese performs less than $50\%$ of \sabd in all three datasets in terms of RR@10.
We argue that when there is a lack of adequate training data, it is less meaningful to compare different deep learning-based models.

Dataset-wise, all approaches perform relatively worse on the Spark dataset and relatively better on the Kibana dataset.
The observation aligns with the findings from prior studies~\cite{rodrigues2020soft}: the same DBRD approach, i.e., \sabd, achieves a variety of RR@10 on different datasets examined, ranging from $0.55$ (on OpenOffice dataset) to $0.7$ (on Netbeans dataset).
The results demonstrate that the performance of a DBRD technique is dependent on the characteristics of the dataset being used. 
This observation suggests that it would be advantageous to tune the prompt template based on the unique characteristics of each dataset, such as incorporating dataset-specific stopwords into the prompts to prevent the LLMs from listing them as keywords. 
Tailoring the prompt templates to the nuances of individual datasets can potentially enhance the effectiveness of the DBRD technique and improve its overall performance.
We leave this for future work to boost the performance further.

\begin{figure}[t] 
  \begin{subfigure}[b]{0.5\linewidth}
    \centering
    \includegraphics[width=0.6\linewidth]{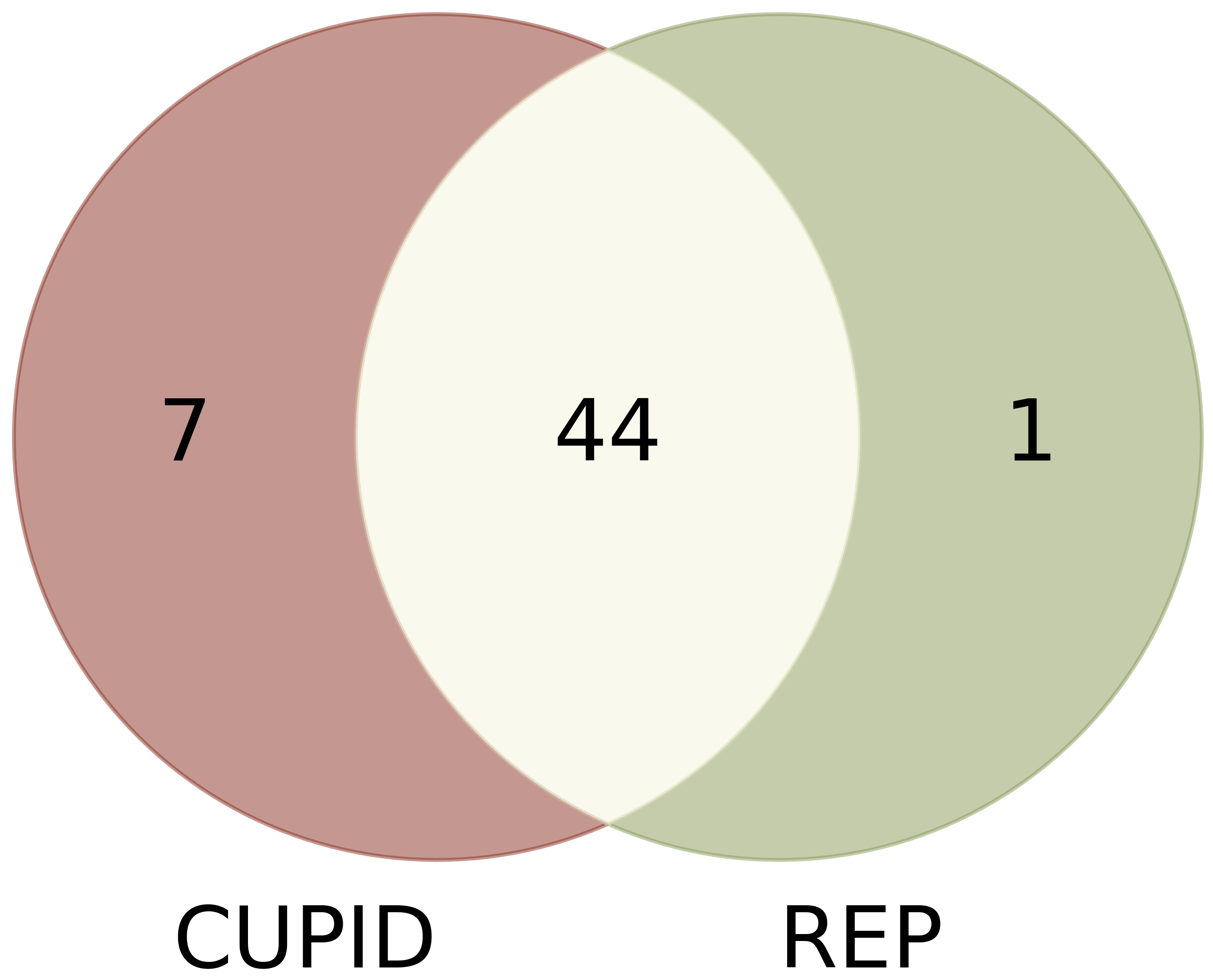} 
    \caption{on Spark dataset} 
    \label{fig7:a} 
    \vspace{4ex}
  \end{subfigure}
  \begin{subfigure}[b]{0.5\linewidth}
    \centering
    \includegraphics[width=0.6\linewidth]{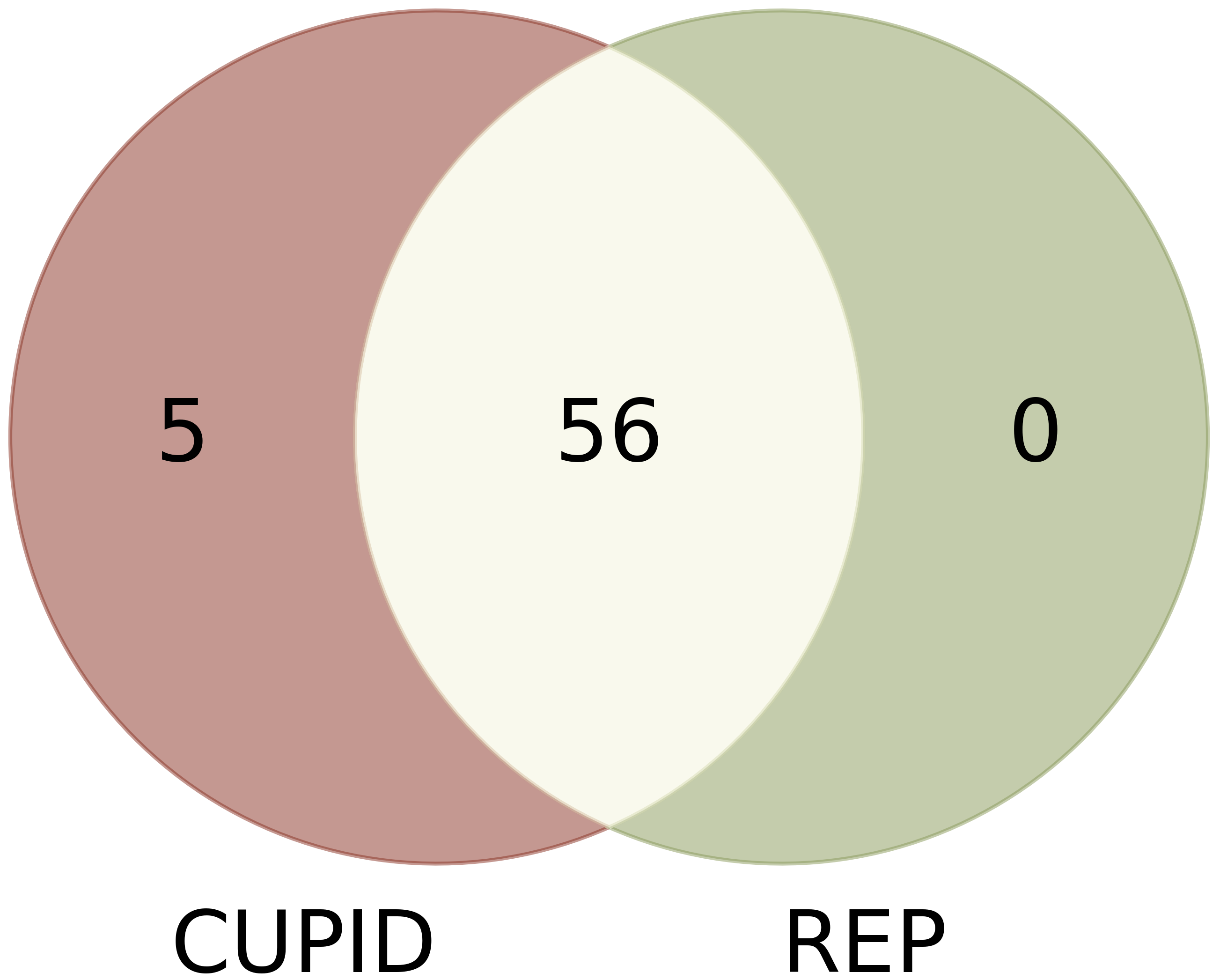} 
    \caption{on Hadoop dataset} 
    \label{fig7:b} 
    \vspace{4ex}
  \end{subfigure} 
  \begin{subfigure}[b]{0.5\linewidth}
    \centering
    \includegraphics[width=0.6\linewidth]{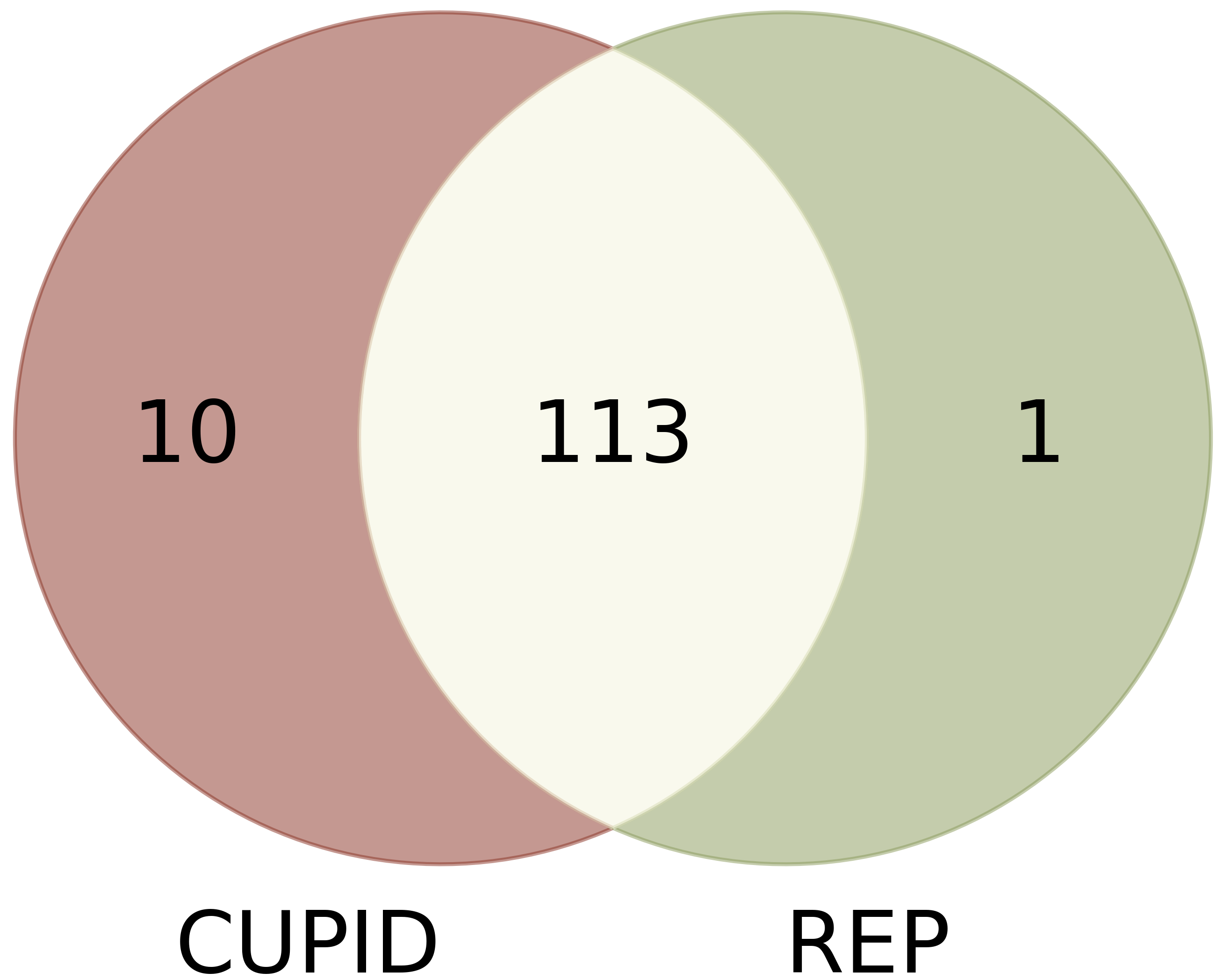} 
    \caption{on Kibana dataset} 
    \label{fig7:c} 
  \end{subfigure}
  \begin{subfigure}[b]{0.5\linewidth}
    \centering
    \includegraphics[width=0.6\linewidth]{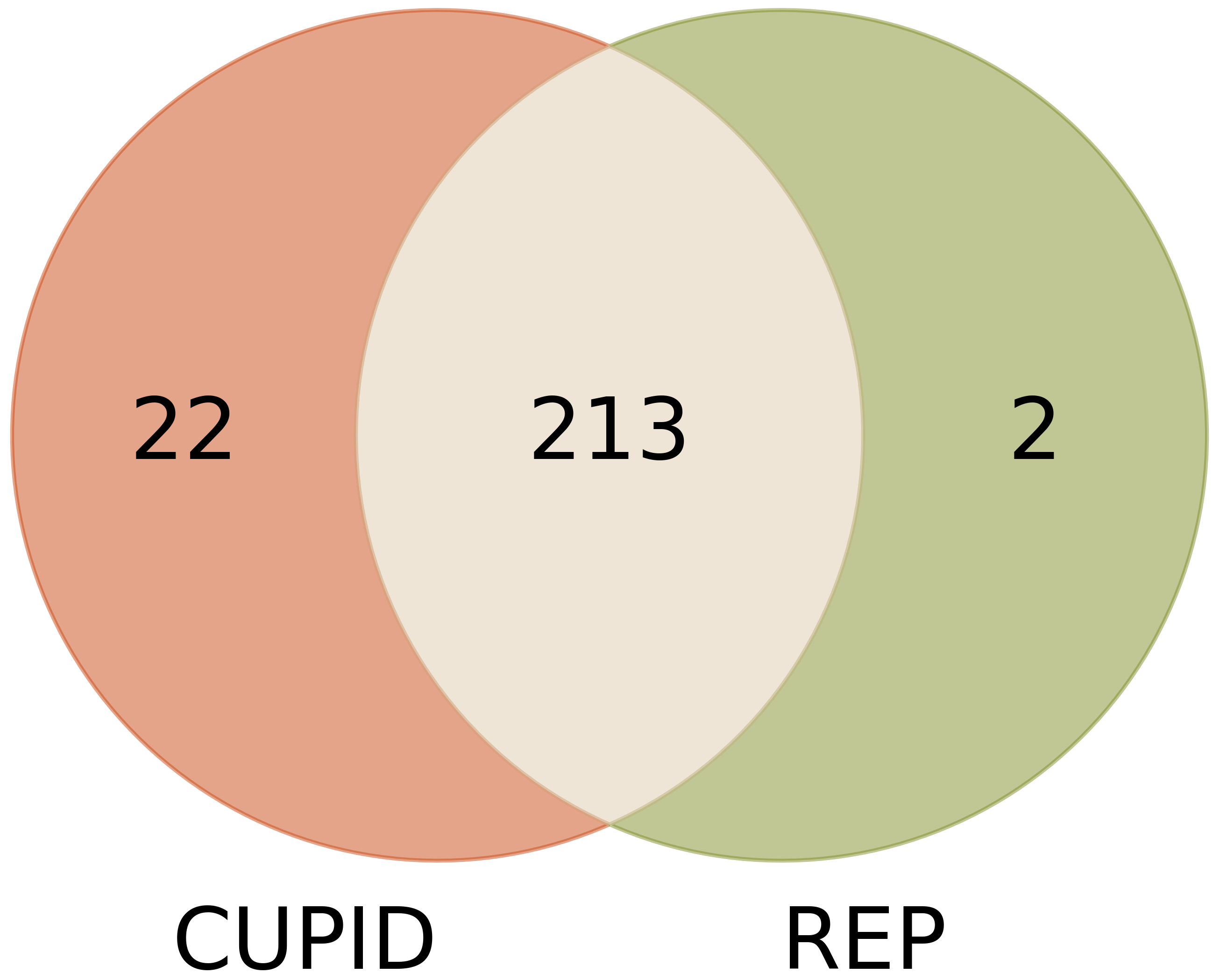} 
    \caption{All the three datasets} 
    \label{fig7:d} 
  \end{subfigure} 
  \caption{Successful prediction Venn diagram}
  \label{fig:venn} 
\end{figure}

Figure~\ref{fig:venn} shows the Venn diagrams for successful predictions made by the prior state-of-the-art method, i.e., \rep, and \method on each dataset and all datasets combined.
We see that \method successfully retrieves more master bug reports compared to \rep.
On the Hadoop dataset, only \method successfully retrieved more master bug reports, while \rep did not successfully retrieve more.

\begin{figure}[t]
\includegraphics[width=\columnwidth]{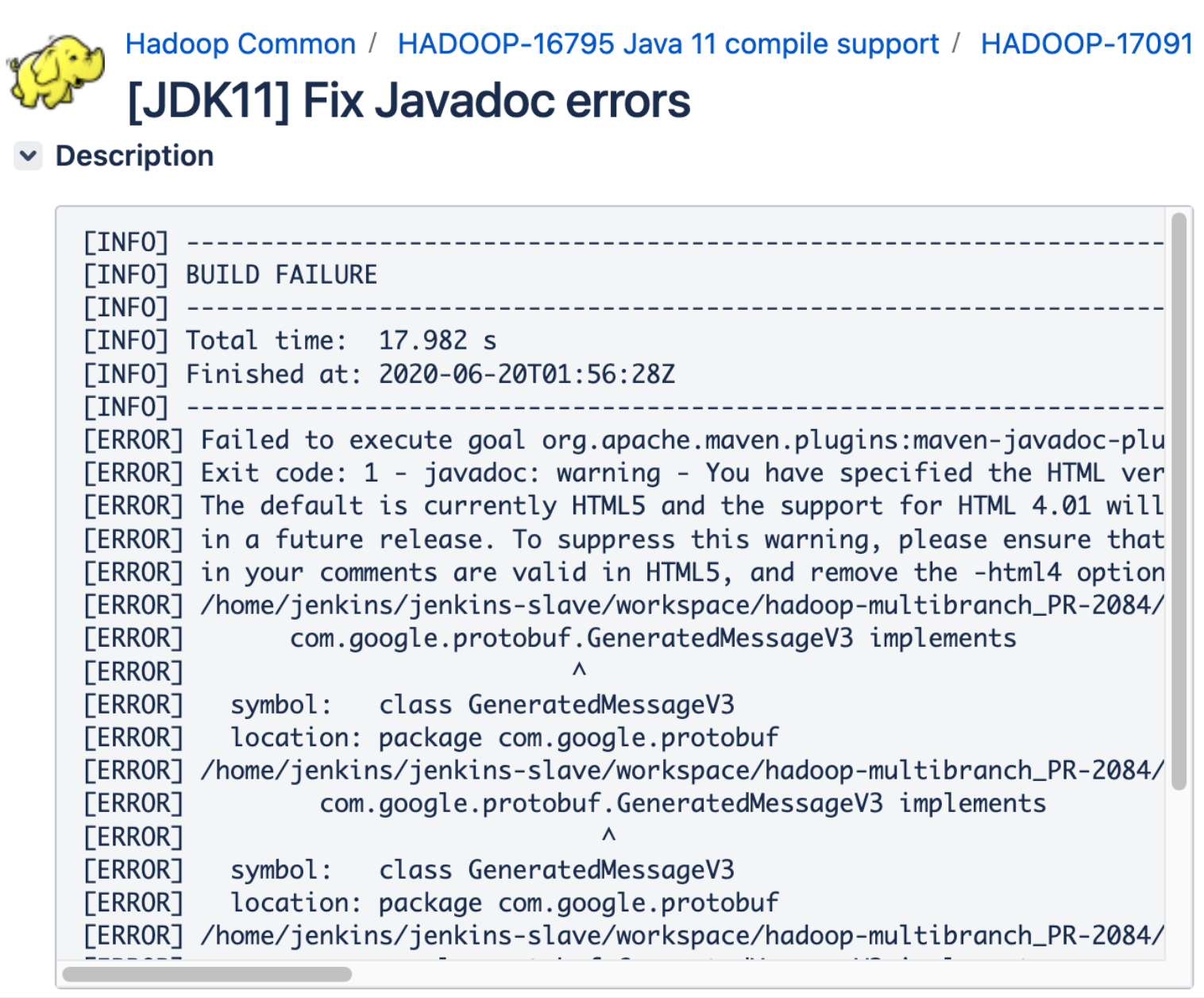}
	\caption{The case where \method succeeded while \rep failed: HADOOP-17091}
	\label{fig:success_cupid}
\end{figure}

To demonstrate the ability of \method, we show an example, i.e., the query bug report is \texttt{HADOOP-17091}\cite{HADOOP17091:online} where \rep failed to predict the correct master bug report in the top-10 positions, while \method managed to.
Figure~\ref{fig:success_cupid} shows the summary and description of this issue.
We can see that there is no natural language in the description, containing only error messages.
Thus, \rep considered the most possible master bug report to be \texttt{HADOOP-16648}~\cite{HADOOP16648:online}, which also contains a large portion of the error messages.
We checked the single-run result by ChatGPT.
Thanks to the language understanding and generation ability of ChatGPT, \method identified the keywords: \texttt{Maven, BUILD FAILURE, Javadoc, error, HTML version, HTML5, HTML4, warning, symbol, class, GeneratedMessageV3, package, com.google.protobuf} from the description of \texttt{HADOOP-17091}.
The generated shorter description on the query bug report has several words that overlap with the description of the real master bug report (\texttt{HADOOP-16862}~\cite{HADOOP16862:online}).
It enables \method to successfully rank it at the top 10 position.
Since the real master bug report has a long error message as the description, \rep failed to retrieve it.
This example shows that ChatGPT can be helpful when long descriptions contain a build failure message.
It can generate the most important keywords, which are vital for duplicate detection.

\textbf{Error Analysis.} We also investigate the two cases when \method failed to predict the correct master bug report in the top-10 positions, while \rep managed to do so. One sample is \texttt{SPARK-31519}\cite{SPARK-31519:online}. 
If we check the keywords extracted by ChatGPT solely, which are \texttt{Spark, SQL parser, Filter operator, Aggregate operator, ResolveAggregateFunctions, ResolveReferences, ResolveFunctions, ResolveTimeZone, Analyzer rule, Grouping columns}, we can see that the keywords are reasonable. However, since the description shrinks the test history, which contains the command and output, \method failed to retrieve the correct master bug report \texttt{SPARK-31334}\cite{SPARK-31334:online}, as it also contains the test history similar to the query bug report. 
The other sample is issue \texttt{\#66249}\cite{kibana66249:online} in Kibana. The query bug report contains the Kibana issue template. The keywords extracted by ChatGPT include the template keywords, which are useless for duplicate detection. Apart from the template keywords, ChatGPT considers \texttt{FireFox} as one of the keywords. This leads to our method retrieving the wrong master bug report \texttt{\#43282}\cite{kibana43282:online} at the first position, which is a bug report related to FireFox. These examples demonstrate that, in some cases, keywords alone are not sufficient to identify duplicate bug reports accurately. 
To address such cases, we can consider using a prompt template that guides ChatGPT in retaining concise code and output history while ignoring the template keywords.

\begin{tcolorbox}[left=4pt,right=4pt,top=2pt,bottom=2pt,boxrule=0.5pt]
\textbf{Answer to RQ1:} 
\method outperforms the best baseline by $8\%$, $5\%$, and $6\%$ in terms of Recall Rate@10 on the Spark, Hadoop, and Kibana datasets, respectively.
\end{tcolorbox}

\subsection{RQ2: Ablation Study}
\begin{table}[t]
  \centering
  \caption{Number of test bugs and bugs that need to run after adopting selection rules.}
  \label{tab:num_bugs}
  \begin{tabular}{lrrr}
  \toprule
  \multirow{2}{*}{\bf Selection Rule} & \multicolumn{3}{c}{\bf Dataset} \\
  \cmidrule{2-4}
  & {\bf Spark} & {\bf Hadoop} & {\bf Kibana} \\
  \midrule
  None & 2,841 & 3,740 & 7,167 \\
  w/ Length & 586 & 903 & 1,883 \\
  w/ Content & 1,466 & 1,537 & 3,048 \\
  w/ Length + Content & 1,583 & 1,845 & 3,791 \\
  \bottomrule
  \end{tabular}
  \end{table}
  
\subsubsection{RQ2.1: Comparison of Different Selection Rules} Table~\ref{tab:num_bugs} shows how many bug reports need to query ChatGPT after adopting (1) no-selection rules, (2) selection by length, (3) selection by content, and (4) selection by both length and content.
If we only use \textit{length} as the selection criteria, we will only need to run ChatGPT on $21\%$,	$24\%$, and	$26\%$ of the original test bug reports in Spark, Hadoop, and Kibana datasets, respectively.
While the computational cost would be reduced, it is essentially a trade-off: we want to achieve both efficiency and accuracy, which can be contradictory in some cases.
We want to take full advantage of ChatGPT with minimal computational costs.
Other than length, we also identify the \textit{content} criteria
If we use \textit{content} as the only selection criteria, we will need to run ChatGPT on $52\%$,	$41\%$, and $43\%$ of the original test bug reports in Spark, Hadoop, and Kibana datasets, respectively.
After adopting both length and content criteria, the bug reports needed to be processed by ChatGPT increased and accounted for $56\%$, $49\%$,	and $53\%$, which still saved more than $40\%$ bug reports from processing.

\begin{table}[t]
\centering
\caption{Ablation study on selection rules: Recall Rate@$k$ obtained on the Spark dataset.}
\label{tab:ablation_study_1}
{
\begin{tabular}{lrrrrrr}
\toprule
{\bf Selection} & {\bf RR@1} & {\bf RR@2} & {\bf RR@3} & {\bf RR@4} & {\bf RR@5} & {\bf RR@10}\\
\midrule
None & 0.365	&0.407&	0.464&	0.489&	{\bf 0.519} &	0.575\\
w/ Length & 0.336	&0.398&	0.422&	0.474&	0.481&	0.568 \\
w/ Content &  {\bf 0.360}	& {\bf 0.435} &	{\bf 0.472}	& {\bf 0.499}&	{\bf 0.519}	&{\bf 0.602} \\
w/ Length + Content & 0.348	&0.422&	0.469	& 0.494 &	0.516	& {\bf 0.602} \\
\bottomrule
\end{tabular}
}
\end{table}

Table~\ref{tab:ablation_study_1} shows the corresponding results of applying selection rules.
Comparing the performance of no selection rules, i.e., querying all the test bug reports with ChatGPT and applying \textit{both} selection rules, we can observe that after applying the rules, RR@10 improves by $8\%$.
We also observe that only using the content as the selection rule can achieve the same performance as using both selection rules; we can conclude that the content is the most important selection rule.
Other than making an improvement on the accuracy, it also frees at least $40\%$ of the bug reports in the test set from querying ChatGPT.
Here, we do not only save the computational cost but also improve accuracy.
Thus, we use the content as the selection rule in \method.

\begin{table}[t]
\centering
\caption{Ablation study on ChatGPT: Recall Rate@$k$ obtained on the Spark dataset.}
\label{tab:ablation_study_2}
{
\begin{tabular}{lrrrrrr}
\toprule
{\bf Model} & {\bf RR@1} & {\bf RR@2} & {\bf RR@3} & {\bf RR@4} & {\bf RR@5} & {\bf RR@10}\\
\midrule
\llama &  0.321 &	0.432 &	0.444 &	0.469 &	0.506 &	{\bf 0.605} \\
\phithree &  0.309	& 0.383	& 0.432 &	0.481 &	0.494 &	0.519 \\
\openchat & 0.321	& 0.358 &	0.383 &	0.432 &	0.457 &	0.531\\
\midrule
\textsc{TFIDF} & 0.383 &	{\bf 0.457} &	{\bf 0.494} &	0.494&	{\bf 0.519} &	0.556\\
\textsc{KP-Miner} & 0.358	&0.444&	0.469&	0.481	&0.481&	0.531 \\
\textsc{YAKE} &  0.321&	0.407	& 0.481 & 	0.481&	0.494&	0.556\\
\midrule
\method & {\bf 0.360}	& 0.435 & 0.472 &	{\bf 0.499}	& {\bf 0.519}	& 0.602  \\
\bottomrule
\end{tabular}
}
\end{table}

\begin{table}[t]
  \centering
  \caption{Average number of keywords extracted by different methods.}
  \label{tab:ablation_study_2_keyword}
  {
  \begin{tabular}{lrr}
  \toprule
  \multirow{2}{*}{\bf Model} & \multicolumn{2}{c}{{\bf Average Keywords}}\\
  & {\bf Summary} & {\bf Description} \\
  \midrule
  \llama &  5.1 & 10.9 \\
  \phithree & 15.7 & 36.3 \\
  \openchat & 10.7 & 21 \\
  \midrule
  \textsc{TFIDF} & 11.3 & 11.6 \\
  \textsc{KP-Miner} & 11.0 & 13.9 \\
  \textsc{YAKE} & 11.7 & 15.2 \\
  \midrule
  \method & 5.3 & 18.0 \\
  \bottomrule
  \end{tabular}
  }
\end{table}

\subsubsection{RQ2.2: Comparison of Different Keyword Extractors} 
Compared with ChatGPT, \llama achieves comparable performance in terms of RR@10, while \phithree and \openchat perform much worse.
The excellent performance of \llama makes it promising to use an open-source LLM as a substitute for ChatGPT for the DBRD task.
Further investigation on when and why open-source LLMs lose to ChatGPT can be put to take full advantage of the latest advancement of LLMs.
In comparison with the statistics-based keyword extraction methods, ChatGPT outperforms \textsc{TFIDF} by $8\%$, \textsc{KP-Miner} by $13\%$, and \textsc{YAKE} by $8\%$ in terms of RR@10.

Table~\ref{tab:ablation_study_2_keyword} shows the average number of keywords extracted by different methods. We can observe that \llama extracts the least number of keywords, while \phithree and \openchat extract the most. To understand why \phithree and \openchat extract more keywords, we examined the generated keywords. We found that \phithree generated description keywords over 1,000 words on 9 occasions. In such cases, it tended to repeat certain words until reaching the output token limit. For instance, the description keywords generated by \phithree for the bug report \texttt{SPARK-30412}~\cite{SPARK30457:online} contained the phrase \texttt{deprecated methods in Java test suites} 246 times. 
This scenario was observed 3 times in \openchat-, once in \llama-, and 7 times in ChatGPT-generated keywords (note that ChatGPT has been run 4 times more than the other LLMs). 
As it is generally expected that keywords should not be repeatable and can be easily made unique, the low ratio of such cases in our experiment does not seem to be a significant issue.
In the future, we can mitigate this generation issue by keeping only the unique set of the extracted keywords.
For the statistics-based keyword extraction methods, we used the default parameters, with the relevant parameter set to 10, i.e., \texttt{get\_n\_best=10}, retrieving the 10 highest-scored candidates as keywords or phrases.

In terms of efficiency, on average, all the experimented methods took less than 5 seconds to make a prediction.
There is no significant difference in terms of efficiency among the methods.
Note that the actual time LLM takes for each specific bug report is different, as it depends on the length and complexity of the bug report.

\begin{table}[t]
  \centering
  \caption{Ablation study on alternative prompt templates: Recall Rate@$k$ obtained on the Spark dataset.}
  \label{tab:ablation_study_3}
  \begin{tabular}{lrrrrrr}
  \toprule
  {\bf Method} & {\bf RR@1} & {\bf RR@2} & {\bf RR@3} & {\bf RR@4} & {\bf RR@5} & {\bf RR@10}\\
  \midrule    
  Concise& 0.346	& 0.415	& {\bf 0.479}	& {\bf 0.501} &	{\bf 0.526} &	0.595 \\
  Verbose & 0.328	& 0.370	& 0.395	& 0.432 &	0.452	& 0.499 \\
  \midrule
  \method & {\bf 0.360} & {\bf 0.435} &	0.472	& 0.499 &	0.519	& {\bf 0.602}\\
  \bottomrule
  \end{tabular}
\end{table}

\subsubsection{RQ2.3: Comparison of Different Prompt Templates} While running ChatGPT with different prompts, we found sometimes they cannot follow the desired output format.
If ChatGPT's response did not provide separate keywords for the summary and description, instead returning a mixed list of keywords, we would repeat the query up to 5 times. If the response continued to return only a list of keywords without distinguishing between summary and description, we would extract these keywords and use them as both the summary and description.
For all 7,330 queries (i.e., 5 runs of 1,466 test samples in the Spark dataset): (1) using the prompt template from \method, ChatGPT successfully provided separate keywords for the summary and description; (2) with the verbose prompt template, ChatGPT produced a mixed list of keywords 1,239 times; (3) with the concise prompt template, ChatGPT produced a mixed list only 39 times. 
This indicates that when using the more verbose prompt template, ChatGPT tends not to maintain the desired output format.

Table~\ref{tab:ablation_study_3} shows the results of querying ChatGPT with the two alternative templates and the template employed in \method. 
We observe that the verbose prompt, which is more comprehensive than the concise prompt, leads to worse performance, being 17\% worse than the final prompt template in terms of RR@10. 
The concise and final templates convey very similar meanings and do not have a significant impact on performance.
These results indicate the significance of prompt engineering.

\begin{tcolorbox}[left=4pt,right=4pt,top=2pt,bottom=2pt,boxrule=0.5pt]
\textbf{Answer to RQ2:} 
Selection rules effectively improve the performance of \method and reduce the computational cost.
Furthermore, \llama can be a promising open-source LLM alternative for extracting keywords.
Suitable prompt templates are crucial for the performance of ChatGPT.
\end{tcolorbox} 

\section{Threats to Validity}
\label{sec:discussion}
\textbf{Threats to Internal Validity.} The main internal threat is whether there is data leakage in ChatGPT and the open-source LLMs.
However, since we do not have access to the training data of LLMs, we cannot verify whether there is data leakage in LLMs.
Even so, since we did not directly use LLMs to compare whether two bug reports are duplicates, instead, we utilize LLMs indirectly, which may not benefit much from memorizing the training data. 
Thus, we believe this threat is minimal.

\textbf{Threats to External Validity.} The primary external threat is the generalizability of our findings.
Our focus in this study is on datasets with a typical number of bug reports, roughly $10k$ issues. 
Therefore, our results may not extend to datasets with a significantly greater number of bug reports, such as those containing tens of thousands of issues. 
Nevertheless, we believe that our findings remain valuable for the majority of projects. This is supported by the fact that in a dataset of 994 high-quality projects from GitHub, each project contains an average of $2k$ issues~\cite{joshi2019rapidrelease}.

\section{Related Work}
\label{sec:related_work}

Except for the works that studied the DBRD task, other automated bug report management tasks have also attracted much research interest~\cite{zhou2012should,su2021reducing,lee2022light,zhang2022automatic}.
In this section, we briefly introduce several studies on automated bug report management tasks, including bug component assignment~\cite{su2021reducing}, developer assignment~\cite{lee2022light}, and issue title generation~\cite{zhang2022itiger}.
Su et al.~\cite{su2021reducing} propose a learning-to-rank framework that leverages the correct bug assignment history to assign the most appropriate component to a bug.
Instead of only using the features from bug reports, their approach also derives rich features from this knowledge graph.
Lee et al.~\cite{lee2022light} focus on assigning a bug report to the most appropriate developer.
They propose a framework (i.e., LBT-P) that applies LLMs, such as RoBERTa~\cite{liu2019roberta}, to extract semantic information.
LBT-P uses knowledge distillation to compress an LLM into a small and fast model.
Additionally, it also introduces knowledge preservation fine-tuning to handle the challenge of catastrophic forgetting~\cite{mccloskey1989catastrophic} of LLM.
Although the issue is not a new concept, the automatic issue title generation task has not been studied until recently~\cite{chen2020stay,zhang2022itiger}.
Zhang et al.~\cite{zhang2022itiger} propose to leverage the state-of-the-art BART~\cite{lewis2020bart} model to generate issue titles for bug reports.
The experimental results show that fine-tuning BART can better generate issue titles than the prior state-of-the-art approach, i.e., iTAPE~\cite{chen2020stay}, based on sequence-to-sequence model~\cite{sutskever2014sequence}.

\section{Conclusion and Future Work}
\label{sec:conclusion}

In this work, we focus on the task of DBRD, especially on projects with a typical number of bug reports.
We investigated how to combine the advantages of both the traditional DBRD approach and LLMs and proposed \method.
\method leverages ChatGPT to identify keywords as the input of the state-of-the-art traditional DBRD approach \rep.
We conduct a comprehensive evaluation on three datasets and compare \method with three baselines.
We also compare ChatGPT with its alternatives, i.e., three open-source LLMs, and three statistical-based keyword extraction techniques.
The experimental results show that \method outperforms the state-of-the-art DBRD techniques regarding Recall Rate@10 on all the datasets.
Particularly, \method achieves high Recall Rate@10 scores ranging from $0.602$ to $0.654$ on all the datasets investigated.

We envision deploying \method in practical settings without introducing significant overhead. The query time for a single bug report using ChatGPT is remarkably efficient, averaging less than 5 seconds. This rapid turnaround time attests to the efficiency of the underlying LLMs and the optimized implementation. 
It ensures that the integration of \method into existing bug-triaging workflows is seamless. In the future, we plan to package \method as a tool for practitioners. Additionally, we are interested in exploring the application of ChatGPT to other bug report management tasks, such as bug component and developer assignment, leveraging its natural language understanding capabilities.

\textbf{Availability.}  Our replication package is publically available at \url{https://github.com/soarsmu/CUPID}.

\backmatter

\bibliography{main}


\begin{thebibliography}{70}
\ifx \bisbn   \undefined \def \bisbn  #1{ISBN #1}\fi
\ifx \binits  \undefined \def \binits#1{#1}\fi
\ifx \bauthor  \undefined \def \bauthor#1{#1}\fi
\ifx \batitle  \undefined \def \batitle#1{#1}\fi
\ifx \bjtitle  \undefined \def \bjtitle#1{#1}\fi
\ifx \bvolume  \undefined \def \bvolume#1{\textbf{#1}}\fi
\ifx \byear  \undefined \def \byear#1{#1}\fi
\ifx \bissue  \undefined \def \bissue#1{#1}\fi
\ifx \bfpage  \undefined \def \bfpage#1{#1}\fi
\ifx \blpage  \undefined \def \blpage #1{#1}\fi
\ifx \burl  \undefined \def \burl#1{\textsf{#1}}\fi
\ifx \doiurl  \undefined \def \doiurl#1{\url{https://doi.org/#1}}\fi
\ifx \betal  \undefined \def \betal{\textit{et al.}}\fi
\ifx \binstitute  \undefined \def \binstitute#1{#1}\fi
\ifx \binstitutionaled  \undefined \def \binstitutionaled#1{#1}\fi
\ifx \bctitle  \undefined \def \bctitle#1{#1}\fi
\ifx \beditor  \undefined \def \beditor#1{#1}\fi
\ifx \bpublisher  \undefined \def \bpublisher#1{#1}\fi
\ifx \bbtitle  \undefined \def \bbtitle#1{#1}\fi
\ifx \bedition  \undefined \def \bedition#1{#1}\fi
\ifx \bseriesno  \undefined \def \bseriesno#1{#1}\fi
\ifx \blocation  \undefined \def \blocation#1{#1}\fi
\ifx \bsertitle  \undefined \def \bsertitle#1{#1}\fi
\ifx \bsnm \undefined \def \bsnm#1{#1}\fi
\ifx \bsuffix \undefined \def \bsuffix#1{#1}\fi
\ifx \bparticle \undefined \def \bparticle#1{#1}\fi
\ifx \barticle \undefined \def \barticle#1{#1}\fi
\bibcommenthead
\ifx \bconfdate \undefined \def \bconfdate #1{#1}\fi
\ifx \botherref \undefined \def \botherref #1{#1}\fi
\ifx \url \undefined \def \url#1{\textsf{#1}}\fi
\ifx \bchapter \undefined \def \bchapter#1{#1}\fi
\ifx \bbook \undefined \def \bbook#1{#1}\fi
\ifx \bcomment \undefined \def \bcomment#1{#1}\fi
\ifx \oauthor \undefined \def \oauthor#1{#1}\fi
\ifx \citeauthoryear \undefined \def \citeauthoryear#1{#1}\fi
\ifx \endbibitem  \undefined \def \endbibitem {}\fi
\ifx \bconflocation  \undefined \def \bconflocation#1{#1}\fi
\ifx \arxivurl  \undefined \def \arxivurl#1{\textsf{#1}}\fi
\csname PreBibitemsHook\endcsname

\bibitem[\protect\citeauthoryear{}{2024}]{Bugzilla40:online}
\begin{botherref}
Bugzilla.
\url{https://www.bugzilla.org/}.
(Accessed on 05/05/2024)
(2024)
\end{botherref}
\endbibitem

\bibitem[\protect\citeauthoryear{}{2024}]{JiraIssu14:online}
\begin{botherref}
Jira | Issue \& Project Tracking Software | Atlassian.
\url{https://www.atlassian.com/software/jira}.
(Accessed on 05/05/2024)
(2024)
\end{botherref}
\endbibitem

\bibitem[\protect\citeauthoryear{}{2024}]{GitHub24:online}
\begin{botherref}
GitHub.
\url{https://github.com/}.
(Accessed on 05/05/2024)
(2024)
\end{botherref}
\endbibitem

\bibitem[\protect\citeauthoryear{Lazar et~al.}{2014}]{lazar2014generating}
\begin{bchapter}
\bauthor{\bsnm{Lazar}, \binits{A.}},
\bauthor{\bsnm{Ritchey}, \binits{S.}},
\bauthor{\bsnm{Sharif}, \binits{B.}}:
\bctitle{Generating duplicate bug datasets}.
In: \bbtitle{Proceedings of the 11th Working Conference on Mining Software Repositories},
pp. \bfpage{392}--\blpage{395}
(\byear{2014})
\end{bchapter}
\endbibitem

\bibitem[\protect\citeauthoryear{Sun et~al.}{2011}]{sun2011towards}
\begin{bchapter}
\bauthor{\bsnm{Sun}, \binits{C.}},
\bauthor{\bsnm{Lo}, \binits{D.}},
\bauthor{\bsnm{Khoo}, \binits{S.-C.}},
\bauthor{\bsnm{Jiang}, \binits{J.}}:
\bctitle{Towards more accurate retrieval of duplicate bug reports}.
In: \bbtitle{2011 26th IEEE/ACM International Conference on Automated Software Engineering (ASE 2011)},
pp. \bfpage{253}--\blpage{262}
(\byear{2011}).
\bcomment{IEEE}
\end{bchapter}
\endbibitem

\bibitem[\protect\citeauthoryear{Deshmukh et~al.}{2017}]{deshmukh2017towards}
\begin{bchapter}
\bauthor{\bsnm{Deshmukh}, \binits{J.}},
\bauthor{\bsnm{Annervaz}, \binits{K.}},
\bauthor{\bsnm{Podder}, \binits{S.}},
\bauthor{\bsnm{Sengupta}, \binits{S.}},
\bauthor{\bsnm{Dubash}, \binits{N.}}:
\bctitle{Towards accurate duplicate bug retrieval using deep learning techniques}.
In: \bbtitle{2017 IEEE International Conference on Software Maintenance and Evolution (ICSME)},
pp. \bfpage{115}--\blpage{124}
(\byear{2017}).
\bcomment{IEEE}
\end{bchapter}
\endbibitem

\bibitem[\protect\citeauthoryear{Rodrigues et~al.}{2020}]{rodrigues2020soft}
\begin{bchapter}
\bauthor{\bsnm{Rodrigues}, \binits{I.M.}},
\bauthor{\bsnm{Aloise}, \binits{D.}},
\bauthor{\bsnm{Fernandes}, \binits{E.R.}},
\bauthor{\bsnm{Dagenais}, \binits{M.}}:
\bctitle{A soft alignment model for bug deduplication}.
In: \bbtitle{Proceedings of the 17th International Conference on Mining Software Repositories},
pp. \bfpage{43}--\blpage{53}
(\byear{2020})
\end{bchapter}
\endbibitem

\bibitem[\protect\citeauthoryear{He et~al.}{2020}]{he2020duplicate}
\begin{bchapter}
\bauthor{\bsnm{He}, \binits{J.}},
\bauthor{\bsnm{Xu}, \binits{L.}},
\bauthor{\bsnm{Yan}, \binits{M.}},
\bauthor{\bsnm{Xia}, \binits{X.}},
\bauthor{\bsnm{Lei}, \binits{Y.}}:
\bctitle{Duplicate bug report detection using dual-channel convolutional neural networks}.
In: \bbtitle{Proceedings of the 28th International Conference on Program Comprehension},
pp. \bfpage{117}--\blpage{127}
(\byear{2020})
\end{bchapter}
\endbibitem

\bibitem[\protect\citeauthoryear{Xiao et~al.}{2020}]{xiao2020hindbr}
\begin{bchapter}
\bauthor{\bsnm{Xiao}, \binits{G.}},
\bauthor{\bsnm{Du}, \binits{X.}},
\bauthor{\bsnm{Sui}, \binits{Y.}},
\bauthor{\bsnm{Yue}, \binits{T.}}:
\bctitle{Hindbr: Heterogeneous information network based duplicate bug report prediction}.
In: \bbtitle{2020 IEEE 31st International Symposium on Software Reliability Engineering (ISSRE)},
pp. \bfpage{195}--\blpage{206}
(\byear{2020}).
\bcomment{IEEE}
\end{bchapter}
\endbibitem

\bibitem[\protect\citeauthoryear{Riazi et~al.}{2019}]{riazi2019deep}
\begin{barticle}
\bauthor{\bsnm{Riazi}, \binits{M.S.}},
\bauthor{\bsnm{Rouani}, \binits{B.D.}},
\bauthor{\bsnm{Koushanfar}, \binits{F.}}:
\batitle{Deep learning on private data}.
\bjtitle{IEEE Security \& Privacy}
\bvolume{17}(\bissue{6}),
\bfpage{54}--\blpage{63}
(\byear{2019})
\end{barticle}
\endbibitem

\bibitem[\protect\citeauthoryear{Joshi and Chimalakonda}{2019}]{joshi2019rapidrelease}
\begin{bchapter}
\bauthor{\bsnm{Joshi}, \binits{S.D.}},
\bauthor{\bsnm{Chimalakonda}, \binits{S.}}:
\bctitle{Rapidrelease-a dataset of projects and issues on github with rapid releases}.
In: \bbtitle{2019 IEEE/ACM 16th International Conference on Mining Software Repositories (MSR)},
pp. \bfpage{587}--\blpage{591}
(\byear{2019}).
\bcomment{IEEE}
\end{bchapter}
\endbibitem

\bibitem[\protect\citeauthoryear{}{2023}]{ohmyzsho48:online}
\begin{botherref}
ohmyzsh/ohmyzsh.
\url{https://github.com/ohmyzsh/ohmyzsh}.
(Accessed on 05/05/2024)
(2023)
\end{botherref}
\endbibitem

\bibitem[\protect\citeauthoryear{}{2024}]{axiosaxi17:online}
\begin{botherref}
axios/axios: Promise based HTTP client for the browser and node.js.
\url{https://github.com/axios/axios}.
(Accessed on 05/05/2024)
(2024)
\end{botherref}
\endbibitem

\bibitem[\protect\citeauthoryear{}{2024}]{vuejsvue14:online}
\begin{botherref}
vuejs/vue: This is the repo for Vue 2. For Vue 3, go to https://github.com/vuejs/core.
\url{https://github.com/vuejs/vue}.
(Accessed on 05/05/2024)
(2024)
\end{botherref}
\endbibitem

\bibitem[\protect\citeauthoryear{}{2024}]{Signific69:online}
\begin{botherref}
Significant-Gravitas/Auto-GPT.
\url{https://github.com/Significant-Gravitas/Auto-GPT}.
(Accessed on 05/05/2024)
(2024)
\end{botherref}
\endbibitem

\bibitem[\protect\citeauthoryear{Zhang et~al.}{2022}]{zhang2022duplicate}
\begin{botherref}
\oauthor{\bsnm{Zhang}, \binits{T.}},
\oauthor{\bsnm{Han}, \binits{D.}},
\oauthor{\bsnm{Vinayakarao}, \binits{V.}},
\oauthor{\bsnm{Irsan}, \binits{I.C.}},
\oauthor{\bsnm{Xu}, \binits{B.}},
\oauthor{\bsnm{Thung}, \binits{F.}},
\oauthor{\bsnm{Lo}, \binits{D.}},
\oauthor{\bsnm{Jiang}, \binits{L.}}:
Duplicate bug report detection: How far are we?
ACM Transactions on Software Engineering and Methodology
(2022)
\end{botherref}
\endbibitem

\bibitem[\protect\citeauthoryear{Runeson et~al.}{2007}]{runeson2007detection}
\begin{bchapter}
\bauthor{\bsnm{Runeson}, \binits{P.}},
\bauthor{\bsnm{Alexandersson}, \binits{M.}},
\bauthor{\bsnm{Nyholm}, \binits{O.}}:
\bctitle{Detection of duplicate defect reports using natural language processing}.
In: \bbtitle{29th International Conference on Software Engineering (ICSE'07)},
pp. \bfpage{499}--\blpage{510}
(\byear{2007}).
\bcomment{IEEE}
\end{bchapter}
\endbibitem

\bibitem[\protect\citeauthoryear{Jalbert and Weimer}{2008}]{jalbert2008automated}
\begin{bchapter}
\bauthor{\bsnm{Jalbert}, \binits{N.}},
\bauthor{\bsnm{Weimer}, \binits{W.}}:
\bctitle{Automated duplicate detection for bug tracking systems}.
In: \bbtitle{2008 IEEE International Conference on Dependable Systems and Networks With FTCS and DCC (DSN)},
pp. \bfpage{52}--\blpage{61}
(\byear{2008}).
\bcomment{IEEE}
\end{bchapter}
\endbibitem

\bibitem[\protect\citeauthoryear{Sun et~al.}{2010}]{sun2010discriminative}
\begin{bchapter}
\bauthor{\bsnm{Sun}, \binits{C.}},
\bauthor{\bsnm{Lo}, \binits{D.}},
\bauthor{\bsnm{Wang}, \binits{X.}},
\bauthor{\bsnm{Jiang}, \binits{J.}},
\bauthor{\bsnm{Khoo}, \binits{S.-C.}}:
\bctitle{A discriminative model approach for accurate duplicate bug report retrieval}.
In: \bbtitle{Proceedings of the 32nd ACM/IEEE International Conference on Software Engineering-Volume 1},
pp. \bfpage{45}--\blpage{54}
(\byear{2010})
\end{bchapter}
\endbibitem

\bibitem[\protect\citeauthoryear{AI@Meta}{2024}]{llama3modelcard}
\begin{botherref}
\oauthor{\bsnm{AI@Meta}}:
Llama 3 model card
(2024)
\end{botherref}
\endbibitem

\bibitem[\protect\citeauthoryear{}{2024}]{phi-blog}
\begin{botherref}
Introducing Phi-3: Redefining what's possible with SLMs | Microsoft Azure Blog.
\url{https://azure.microsoft.com/en-us/blog/introducing-phi-3-redefining-whats-possible-with-slms/}.
(Accessed on 05/13/2024)
(2024)
\end{botherref}
\endbibitem

\bibitem[\protect\citeauthoryear{Abdin et~al.}{2024}]{abdin2024phi}
\begin{botherref}
\oauthor{\bsnm{Abdin}, \binits{M.}},
\oauthor{\bsnm{Jacobs}, \binits{S.A.}},
\oauthor{\bsnm{Awan}, \binits{A.A.}},
\oauthor{\bsnm{Aneja}, \binits{J.}},
\oauthor{\bsnm{Awadallah}, \binits{A.}},
\oauthor{\bsnm{Awadalla}, \binits{H.}},
\oauthor{\bsnm{Bach}, \binits{N.}},
\oauthor{\bsnm{Bahree}, \binits{A.}},
\oauthor{\bsnm{Bakhtiari}, \binits{A.}},
\oauthor{\bsnm{Behl}, \binits{H.}}, et al.:
Phi-3 technical report: A highly capable language model locally on your phone.
arXiv preprint arXiv:2404.14219
(2024)
\end{botherref}
\endbibitem

\bibitem[\protect\citeauthoryear{Wang et~al.}{2024}]{wang2023openchat}
\begin{bchapter}
\bauthor{\bsnm{Wang}, \binits{G.}},
\bauthor{\bsnm{Cheng}, \binits{S.}},
\bauthor{\bsnm{Zhan}, \binits{X.}},
\bauthor{\bsnm{Li}, \binits{X.}},
\bauthor{\bsnm{Song}, \binits{S.}},
\bauthor{\bsnm{Liu}, \binits{Y.}}:
\bctitle{Openchat: Advancing open-source language models with mixed-quality data}.
In: \bbtitle{The Twelfth International Conference on Learning Representations}
(\byear{2024})
\end{bchapter}
\endbibitem

\bibitem[\protect\citeauthoryear{Devlin et~al.}{2019}]{devlin2019bert}
\begin{bchapter}
\bauthor{\bsnm{Devlin}, \binits{J.}},
\bauthor{\bsnm{Chang}, \binits{M.-W.}},
\bauthor{\bsnm{Lee}, \binits{K.}},
\bauthor{\bsnm{Toutanova}, \binits{K.}}:
\bctitle{Bert: Pre-training of deep bidirectional transformers for language understanding}.
In: \bbtitle{Proceedings of NAACL-HLT},
pp. \bfpage{4171}--\blpage{4186}
(\byear{2019})
\end{bchapter}
\endbibitem

\bibitem[\protect\citeauthoryear{Brown et~al.}{2020}]{brown2020language}
\begin{barticle}
\bauthor{\bsnm{Brown}, \binits{T.}},
\bauthor{\bsnm{Mann}, \binits{B.}},
\bauthor{\bsnm{Ryder}, \binits{N.}},
\bauthor{\bsnm{Subbiah}, \binits{M.}},
\bauthor{\bsnm{Kaplan}, \binits{J.D.}},
\bauthor{\bsnm{Dhariwal}, \binits{P.}},
\bauthor{\bsnm{Neelakantan}, \binits{A.}},
\bauthor{\bsnm{Shyam}, \binits{P.}},
\bauthor{\bsnm{Sastry}, \binits{G.}},
\bauthor{\bsnm{Askell}, \binits{A.}}, \betal:
\batitle{Language models are few-shot learners}.
\bjtitle{Advances in neural information processing systems}
\bvolume{33},
\bfpage{1877}--\blpage{1901}
(\byear{2020})
\end{barticle}
\endbibitem

\bibitem[\protect\citeauthoryear{Radford et~al.}{2019}]{radford2019language}
\begin{barticle}
\bauthor{\bsnm{Radford}, \binits{A.}},
\bauthor{\bsnm{Wu}, \binits{J.}},
\bauthor{\bsnm{Child}, \binits{R.}},
\bauthor{\bsnm{Luan}, \binits{D.}},
\bauthor{\bsnm{Amodei}, \binits{D.}},
\bauthor{\bsnm{Sutskever}, \binits{I.}}, \betal:
\batitle{Language models are unsupervised multitask learners}.
\bjtitle{OpenAI blog}
\bvolume{1}(\bissue{8}),
\bfpage{9}
(\byear{2019})
\end{barticle}
\endbibitem

\bibitem[\protect\citeauthoryear{}{2024}]{chatgpt:online}
\begin{botherref}
Introducing ChatGPT.
\url{https://openai.com/blog/chatgpt}.
(Accessed on 05/05/2024)
(2024)
\end{botherref}
\endbibitem

\bibitem[\protect\citeauthoryear{Bettenburg et~al.}{2008}]{bettenburg2008makes}
\begin{bchapter}
\bauthor{\bsnm{Bettenburg}, \binits{N.}},
\bauthor{\bsnm{Just}, \binits{S.}},
\bauthor{\bsnm{Schr{\"o}ter}, \binits{A.}},
\bauthor{\bsnm{Weiss}, \binits{C.}},
\bauthor{\bsnm{Premraj}, \binits{R.}},
\bauthor{\bsnm{Zimmermann}, \binits{T.}}:
\bctitle{What makes a good bug report?}
In: \bbtitle{Proceedings of the 16th ACM SIGSOFT International Symposium on Foundations of Software Engineering},
pp. \bfpage{308}--\blpage{318}
(\byear{2008})
\end{bchapter}
\endbibitem

\bibitem[\protect\citeauthoryear{Wang et~al.}{2008}]{wang2008approach}
\begin{bchapter}
\bauthor{\bsnm{Wang}, \binits{X.}},
\bauthor{\bsnm{Zhang}, \binits{L.}},
\bauthor{\bsnm{Xie}, \binits{T.}},
\bauthor{\bsnm{Anvik}, \binits{J.}},
\bauthor{\bsnm{Sun}, \binits{J.}}:
\bctitle{An approach to detecting duplicate bug reports using natural language and execution information}.
In: \bbtitle{Proceedings of the 30th International Conference on Software Engineering},
pp. \bfpage{461}--\blpage{470}
(\byear{2008})
\end{bchapter}
\endbibitem

\bibitem[\protect\citeauthoryear{Nguyen et~al.}{2012}]{nguyen2012duplicate}
\begin{bchapter}
\bauthor{\bsnm{Nguyen}, \binits{A.T.}},
\bauthor{\bsnm{Nguyen}, \binits{T.T.}},
\bauthor{\bsnm{Nguyen}, \binits{T.N.}},
\bauthor{\bsnm{Lo}, \binits{D.}},
\bauthor{\bsnm{Sun}, \binits{C.}}:
\bctitle{Duplicate bug report detection with a combination of information retrieval and topic modeling}.
In: \bbtitle{Proceedings of the 27th IEEE/ACM International Conference on Automated Software Engineering},
pp. \bfpage{70}--\blpage{79}
(\byear{2012})
\end{bchapter}
\endbibitem

\bibitem[\protect\citeauthoryear{Sureka and Jalote}{2010}]{sureka2010detecting}
\begin{bchapter}
\bauthor{\bsnm{Sureka}, \binits{A.}},
\bauthor{\bsnm{Jalote}, \binits{P.}}:
\bctitle{Detecting duplicate bug report using character n-gram-based features}.
In: \bbtitle{2010 Asia Pacific Software Engineering Conference},
pp. \bfpage{366}--\blpage{374}
(\byear{2010}).
\bcomment{IEEE}
\end{bchapter}
\endbibitem

\bibitem[\protect\citeauthoryear{Pennington et~al.}{2014}]{pennington2014glove}
\begin{bchapter}
\bauthor{\bsnm{Pennington}, \binits{J.}},
\bauthor{\bsnm{Socher}, \binits{R.}},
\bauthor{\bsnm{Manning}, \binits{C.D.}}:
\bctitle{Glove: Global vectors for word representation}.
In: \bbtitle{Proceedings of the 2014 Conference on Empirical Methods in Natural Language Processing (EMNLP)},
pp. \bfpage{1532}--\blpage{1543}
(\byear{2014})
\end{bchapter}
\endbibitem

\bibitem[\protect\citeauthoryear{Mikolov et~al.}{2013}]{mikolov2013efficient}
\begin{botherref}
\oauthor{\bsnm{Mikolov}, \binits{T.}},
\oauthor{\bsnm{Chen}, \binits{K.}},
\oauthor{\bsnm{Corrado}, \binits{G.}},
\oauthor{\bsnm{Dean}, \binits{J.}}:
Efficient estimation of word representations in vector space.
arXiv preprint arXiv:1301.3781
(2013)
\end{botherref}
\endbibitem

\bibitem[\protect\citeauthoryear{Fu et~al.}{2017}]{fu2017hin2vec}
\begin{bchapter}
\bauthor{\bsnm{Fu}, \binits{T.-y.}},
\bauthor{\bsnm{Lee}, \binits{W.-C.}},
\bauthor{\bsnm{Lei}, \binits{Z.}}:
\bctitle{Hin2vec: Explore meta-paths in heterogeneous information networks for representation learning}.
In: \bbtitle{Proceedings of the 2017 ACM on Conference on Information and Knowledge Management},
pp. \bfpage{1797}--\blpage{1806}
(\byear{2017})
\end{bchapter}
\endbibitem

\bibitem[\protect\citeauthoryear{Liu et~al.}{2023}]{liu2023pre}
\begin{barticle}
\bauthor{\bsnm{Liu}, \binits{P.}},
\bauthor{\bsnm{Yuan}, \binits{W.}},
\bauthor{\bsnm{Fu}, \binits{J.}},
\bauthor{\bsnm{Jiang}, \binits{Z.}},
\bauthor{\bsnm{Hayashi}, \binits{H.}},
\bauthor{\bsnm{Neubig}, \binits{G.}}:
\batitle{Pre-train, prompt, and predict: A systematic survey of prompting methods in natural language processing}.
\bjtitle{ACM Computing Surveys}
\bvolume{55}(\bissue{9}),
\bfpage{1}--\blpage{35}
(\byear{2023})
\end{barticle}
\endbibitem

\bibitem[\protect\citeauthoryear{Bang et~al.}{2023}]{bang2023multitask}
\begin{botherref}
\oauthor{\bsnm{Bang}, \binits{Y.}},
\oauthor{\bsnm{Cahyawijaya}, \binits{S.}},
\oauthor{\bsnm{Lee}, \binits{N.}},
\oauthor{\bsnm{Dai}, \binits{W.}},
\oauthor{\bsnm{Su}, \binits{D.}},
\oauthor{\bsnm{Wilie}, \binits{B.}},
\oauthor{\bsnm{Lovenia}, \binits{H.}},
\oauthor{\bsnm{Ji}, \binits{Z.}},
\oauthor{\bsnm{Yu}, \binits{T.}},
\oauthor{\bsnm{Chung}, \binits{W.}}, et al.:
A multitask, multilingual, multimodal evaluation of chatgpt on reasoning, hallucination, and interactivity.
arXiv preprint arXiv:2302.04023
(2023)
\end{botherref}
\endbibitem

\bibitem[\protect\citeauthoryear{Zhang et~al.}{2023a}]{zhang2023revisiting}
\begin{botherref}
\oauthor{\bsnm{Zhang}, \binits{T.}},
\oauthor{\bsnm{Irsan}, \binits{I.C.}},
\oauthor{\bsnm{Thung}, \binits{F.}},
\oauthor{\bsnm{Lo}, \binits{D.}}:
Revisiting sentiment analysis for software engineering in the era of large language models.
arXiv preprint arXiv:2310.11113
(2023)
\end{botherref}
\endbibitem

\bibitem[\protect\citeauthoryear{Zhang et~al.}{2023b}]{zhang2023sentiment}
\begin{botherref}
\oauthor{\bsnm{Zhang}, \binits{W.}},
\oauthor{\bsnm{Deng}, \binits{Y.}},
\oauthor{\bsnm{Liu}, \binits{B.}},
\oauthor{\bsnm{Pan}, \binits{S.J.}},
\oauthor{\bsnm{Bing}, \binits{L.}}:
Sentiment analysis in the era of large language models: A reality check.
arXiv preprint arXiv:2305.15005
(2023)
\end{botherref}
\endbibitem

\bibitem[\protect\citeauthoryear{Robertson et~al.}{2004}]{robertson2004simple}
\begin{bchapter}
\bauthor{\bsnm{Robertson}, \binits{S.}},
\bauthor{\bsnm{Zaragoza}, \binits{H.}},
\bauthor{\bsnm{Taylor}, \binits{M.}}:
\bctitle{Simple bm25 extension to multiple weighted fields}.
In: \bbtitle{Proceedings of the Thirteenth ACM International Conference on Information and Knowledge Management},
pp. \bfpage{42}--\blpage{49}
(\byear{2004})
\end{bchapter}
\endbibitem

\bibitem[\protect\citeauthoryear{Zaragoza et~al.}{2004}]{zaragoza2004microsoft}
\begin{bchapter}
\bauthor{\bsnm{Zaragoza}, \binits{H.}},
\bauthor{\bsnm{Craswell}, \binits{N.}},
\bauthor{\bsnm{Taylor}, \binits{M.J.}},
\bauthor{\bsnm{Saria}, \binits{S.}},
\bauthor{\bsnm{Robertson}, \binits{S.E.}}:
\bctitle{Microsoft cambridge at trec 13: Web and hard tracks.}
In: \bbtitle{Trec},
vol. \bseriesno{4},
pp. \bfpage{1}--\blpage{1}
(\byear{2004})
\end{bchapter}
\endbibitem

\bibitem[\protect\citeauthoryear{Wolf et~al.}{2020}]{wolf-etal-2020-transformers}
\begin{bchapter}
\bauthor{\bsnm{Wolf}, \binits{T.}},
\bauthor{\bsnm{Debut}, \binits{L.}},
\bauthor{\bsnm{Sanh}, \binits{V.}},
\bauthor{\bsnm{Chaumond}, \binits{J.}},
\bauthor{\bsnm{Delangue}, \binits{C.}},
\bauthor{\bsnm{Moi}, \binits{A.}},
\bauthor{\bsnm{Cistac}, \binits{P.}},
\bauthor{\bsnm{Rault}, \binits{T.}},
\bauthor{\bsnm{Louf}, \binits{R.}},
\bauthor{\bsnm{Funtowicz}, \binits{M.}},
\bauthor{\bsnm{Davison}, \binits{J.}},
\bauthor{\bsnm{Shleifer}, \binits{S.}},
\bauthor{\bsnm{Platen}, \binits{P.}},
\bauthor{\bsnm{Ma}, \binits{C.}},
\bauthor{\bsnm{Jernite}, \binits{Y.}},
\bauthor{\bsnm{Plu}, \binits{J.}},
\bauthor{\bsnm{Xu}, \binits{C.}},
\bauthor{\bsnm{Scao}, \binits{T.L.}},
\bauthor{\bsnm{Gugger}, \binits{S.}},
\bauthor{\bsnm{Drame}, \binits{M.}},
\bauthor{\bsnm{Lhoest}, \binits{Q.}},
\bauthor{\bsnm{Rush}, \binits{A.M.}}:
\bctitle{Transformers: State-of-the-art natural language processing}.
In: \bbtitle{Proceedings of the 2020 Conference on Empirical Methods in Natural Language Processing: System Demonstrations},
pp. \bfpage{38}--\blpage{45}.
\bpublisher{Association for Computational Linguistics},
\blocation{Online}
(\byear{2020}).
\burl{https://www.aclweb.org/anthology/2020.emnlp-demos.6}
\end{bchapter}
\endbibitem

\bibitem[\protect\citeauthoryear{}{2024}]{LMSysCha85:online}
\begin{botherref}
LMSys Chatbot Arena Leaderboard - a Hugging Face Space by lmsys.
\url{https://huggingface.co/spaces/lmsys/chatbot-arena-leaderboard}.
(Accessed on 05/05/2024)
(2024)
\end{botherref}
\endbibitem

\bibitem[\protect\citeauthoryear{Rose et~al.}{2010}]{rose2010automatic}
\begin{botherref}
\oauthor{\bsnm{Rose}, \binits{S.}},
\oauthor{\bsnm{Engel}, \binits{D.}},
\oauthor{\bsnm{Cramer}, \binits{N.}},
\oauthor{\bsnm{Cowley}, \binits{W.}}:
Automatic keyword extraction from individual documents.
Text mining: applications and theory,
1--20
(2010)
\end{botherref}
\endbibitem

\bibitem[\protect\citeauthoryear{El-Beltagy and Rafea}{2009}]{el2009kp}
\begin{barticle}
\bauthor{\bsnm{El-Beltagy}, \binits{S.R.}},
\bauthor{\bsnm{Rafea}, \binits{A.}}:
\batitle{Kp-miner: A keyphrase extraction system for english and arabic documents}.
\bjtitle{Information systems}
\bvolume{34}(\bissue{1}),
\bfpage{132}--\blpage{144}
(\byear{2009})
\end{barticle}
\endbibitem

\bibitem[\protect\citeauthoryear{Campos et~al.}{2020}]{campos2020yake}
\begin{barticle}
\bauthor{\bsnm{Campos}, \binits{R.}},
\bauthor{\bsnm{Mangaravite}, \binits{V.}},
\bauthor{\bsnm{Pasquali}, \binits{A.}},
\bauthor{\bsnm{Jorge}, \binits{A.}},
\bauthor{\bsnm{Nunes}, \binits{C.}},
\bauthor{\bsnm{Jatowt}, \binits{A.}}:
\batitle{Yake! keyword extraction from single documents using multiple local features}.
\bjtitle{Information Sciences}
\bvolume{509},
\bfpage{257}--\blpage{289}
(\byear{2020})
\end{barticle}
\endbibitem

\bibitem[\protect\citeauthoryear{Papagiannopoulou and Tsoumakas}{2020}]{papagiannopoulou2020review}
\begin{barticle}
\bauthor{\bsnm{Papagiannopoulou}, \binits{E.}},
\bauthor{\bsnm{Tsoumakas}, \binits{G.}}:
\batitle{A review of keyphrase extraction}.
\bjtitle{Wiley Interdisciplinary Reviews: Data Mining and Knowledge Discovery}
\bvolume{10}(\bissue{2}),
\bfpage{1339}
(\byear{2020})
\end{barticle}
\endbibitem

\bibitem[\protect\citeauthoryear{Amoui et~al.}{2013}]{amoui2013search}
\begin{bchapter}
\bauthor{\bsnm{Amoui}, \binits{M.}},
\bauthor{\bsnm{Kaushik}, \binits{N.}},
\bauthor{\bsnm{Al-Dabbagh}, \binits{A.}},
\bauthor{\bsnm{Tahvildari}, \binits{L.}},
\bauthor{\bsnm{Li}, \binits{S.}},
\bauthor{\bsnm{Liu}, \binits{W.}}:
\bctitle{Search-based duplicate defect detection: An industrial experience}.
In: \bbtitle{2013 10th Working Conference on Mining Software Repositories (MSR)},
pp. \bfpage{173}--\blpage{182}
(\byear{2013}).
\bcomment{IEEE}
\end{bchapter}
\endbibitem

\bibitem[\protect\citeauthoryear{Kochhar et~al.}{2016}]{kochhar2016practitioners}
\begin{bchapter}
\bauthor{\bsnm{Kochhar}, \binits{P.S.}},
\bauthor{\bsnm{Xia}, \binits{X.}},
\bauthor{\bsnm{Lo}, \binits{D.}},
\bauthor{\bsnm{Li}, \binits{S.}}:
\bctitle{Practitioners' expectations on automated fault localization}.
In: \bbtitle{Proceedings of the 25th International Symposium on Software Testing and Analysis},
pp. \bfpage{165}--\blpage{176}
(\byear{2016})
\end{bchapter}
\endbibitem

\bibitem[\protect\citeauthoryear{Sch{\"u}tze et~al.}{2008}]{schutze2008introduction}
\begin{bbook}
\bauthor{\bsnm{Sch{\"u}tze}, \binits{H.}},
\bauthor{\bsnm{Manning}, \binits{C.D.}},
\bauthor{\bsnm{Raghavan}, \binits{P.}}:
\bbtitle{Introduction to Information Retrieval}
vol. \bseriesno{39}.
\bpublisher{Cambridge University Press Cambridge}, \blocation{???}
(\byear{2008})
\end{bbook}
\endbibitem

\bibitem[\protect\citeauthoryear{Parikh et~al.}{2016}]{parikh2016decomposable}
\begin{bchapter}
\bauthor{\bsnm{Parikh}, \binits{A.}},
\bauthor{\bsnm{T{\"a}ckstr{\"o}m}, \binits{O.}},
\bauthor{\bsnm{Das}, \binits{D.}},
\bauthor{\bsnm{Uszkoreit}, \binits{J.}}:
\bctitle{A decomposable attention model for natural language inference}.
In: \bbtitle{Proceedings of the 2016 Conference on Empirical Methods in Natural Language Processing},
pp. \bfpage{2249}--\blpage{2255}
(\byear{2016})
\end{bchapter}
\endbibitem

\bibitem[\protect\citeauthoryear{}{2023}]{ChatGPT4:online}
\begin{botherref}
ChatGPT — Release Notes | OpenAI Help Center.
\url{https://help.openai.com/en/articles/6825453-chatgpt-release-notes}.
(Accessed on 05/05/2024)
(2023)
\end{botherref}
\endbibitem

\bibitem[\protect\citeauthoryear{Boudin}{2016}]{boudin:2016:COLINGDEMO}
\begin{bchapter}
\bauthor{\bsnm{Boudin}, \binits{F.}}:
\bctitle{pke: an open source python-based keyphrase extraction toolkit}.
In: \bbtitle{Proceedings of COLING 2016, the 26th International Conference on Computational Linguistics: System Demonstrations},
\bconflocation{Osaka, Japan},
pp. \bfpage{69}--\blpage{73}
(\byear{2016}).
\burl{http://aclweb.org/anthology/C16-2015}
\end{bchapter}
\endbibitem

\bibitem[\protect\citeauthoryear{}{2024a}]{HADOOP17091:online}
\begin{botherref}
[HADOOP-17091] [JDK11] Fix Javadoc errors - ASF JIRA.
\url{https://issues.apache.org/jira/browse/HADOOP-17091}.
(Accessed on 05/05/2024)
(2024)
\end{botherref}
\endbibitem

\bibitem[\protect\citeauthoryear{}{2024b}]{HADOOP16648:online}
\begin{botherref}
[HADOOP-16648] HDFS Native Client does not build correctly - ASF JIRA.
\url{https://issues.apache.org/jira/browse/HADOOP-16648}.
(Accessed on 05/05/2024)
(2024)
\end{botherref}
\endbibitem

\bibitem[\protect\citeauthoryear{}{2024c}]{HADOOP16862:online}
\begin{botherref}
[HADOOP-16862] [JDK11] Support JavaDoc - ASF JIRA.
\url{https://issues.apache.org/jira/browse/HADOOP-16862}.
(Accessed on 05/05/2024)
(2024)
\end{botherref}
\endbibitem

\bibitem[\protect\citeauthoryear{}{2024a}]{SPARK-31519:online}
\begin{botherref}
[SPARK-31519] Cast in having aggregate expressions returns the wrong result.
\url{https://issues.apache.org/jira/browse/SPARK-31519}.
(Accessed on 05/05/2024)
(2024)
\end{botherref}
\endbibitem

\bibitem[\protect\citeauthoryear{}{2024b}]{SPARK-31334:online}
\begin{botherref}
[SPARK-31334] Use agg column in Having clause behave different with column type.
\url{https://issues.apache.org/jira/browse/SPARK-31334}.
(Accessed on 05/05/2024)
(2024)
\end{botherref}
\endbibitem

\bibitem[\protect\citeauthoryear{}{2024}]{kibana66249:online}
\begin{botherref}
[Discover] "Reset search" is broken · Issue \#66249 · elastic/kibana.
\url{https://github.com/elastic/kibana/issues/66249}.
(Accessed on 05/05/2024)
(2024)
\end{botherref}
\endbibitem

\bibitem[\protect\citeauthoryear{}{}]{kibana43282:online}
\begin{botherref}
FireFox Filters Bug · Issue \#43282 · elastic/kibana.
\url{https://github.com/elastic/kibana/issues/43282}.
(Accessed on 05/05/2024)
\end{botherref}
\endbibitem

\bibitem[\protect\citeauthoryear{}{2024}]{SPARK30457:online}
\begin{botherref}
[SPARK-30412] Eliminate warnings in Java tests regarding to deprecated API - ASF JIRA.
\url{https://issues.apache.org/jira/browse/SPARK-30412}.
(Accessed on 05/05/2024)
(2024)
\end{botherref}
\endbibitem

\bibitem[\protect\citeauthoryear{Zhou et~al.}{2012}]{zhou2012should}
\begin{bchapter}
\bauthor{\bsnm{Zhou}, \binits{J.}},
\bauthor{\bsnm{Zhang}, \binits{H.}},
\bauthor{\bsnm{Lo}, \binits{D.}}:
\bctitle{Where should the bugs be fixed? more accurate information retrieval-based bug localization based on bug reports}.
In: \bbtitle{2012 34th International Conference on Software Engineering (ICSE)},
pp. \bfpage{14}--\blpage{24}
(\byear{2012}).
\bcomment{IEEE}
\end{bchapter}
\endbibitem

\bibitem[\protect\citeauthoryear{Su et~al.}{2021}]{su2021reducing}
\begin{bchapter}
\bauthor{\bsnm{Su}, \binits{Y.}},
\bauthor{\bsnm{Xing}, \binits{Z.}},
\bauthor{\bsnm{Peng}, \binits{X.}},
\bauthor{\bsnm{Xia}, \binits{X.}},
\bauthor{\bsnm{Wang}, \binits{C.}},
\bauthor{\bsnm{Xu}, \binits{X.}},
\bauthor{\bsnm{Zhu}, \binits{L.}}:
\bctitle{Reducing bug triaging confusion by learning from mistakes with a bug tossing knowledge graph}.
In: \bbtitle{2021 36th IEEE/ACM International Conference on Automated Software Engineering (ASE)},
pp. \bfpage{191}--\blpage{202}
(\byear{2021}).
\bcomment{IEEE}
\end{bchapter}
\endbibitem

\bibitem[\protect\citeauthoryear{Lee et~al.}{2022}]{lee2022light}
\begin{bchapter}
\bauthor{\bsnm{Lee}, \binits{J.}},
\bauthor{\bsnm{Han}, \binits{K.}},
\bauthor{\bsnm{Yu}, \binits{H.}}:
\bctitle{A light bug triage framework for applying large pre-trained language model}.
In: \bbtitle{37th IEEE/ACM International Conference on Automated Software Engineering},
pp. \bfpage{1}--\blpage{11}
(\byear{2022})
\end{bchapter}
\endbibitem

\bibitem[\protect\citeauthoryear{Zhang et~al.}{2022a}]{zhang2022automatic}
\begin{bchapter}
\bauthor{\bsnm{Zhang}, \binits{T.}},
\bauthor{\bsnm{Irsan}, \binits{I.C.}},
\bauthor{\bsnm{Thung}, \binits{F.}},
\bauthor{\bsnm{Han}, \binits{D.}},
\bauthor{\bsnm{Lo}, \binits{D.}},
\bauthor{\bsnm{Jiang}, \binits{L.}}:
\bctitle{Automatic pull request title generation}.
In: \bbtitle{2022 IEEE International Conference on Software Maintenance and Evolution (ICSME)},
pp. \bfpage{71}--\blpage{81}
(\byear{2022}).
\bcomment{IEEE Computer Society}
\end{bchapter}
\endbibitem

\bibitem[\protect\citeauthoryear{Zhang et~al.}{2022b}]{zhang2022itiger}
\begin{bchapter}
\bauthor{\bsnm{Zhang}, \binits{T.}},
\bauthor{\bsnm{Irsan}, \binits{I.C.}},
\bauthor{\bsnm{Thung}, \binits{F.}},
\bauthor{\bsnm{Han}, \binits{D.}},
\bauthor{\bsnm{Lo}, \binits{D.}},
\bauthor{\bsnm{Jiang}, \binits{L.}}:
\bctitle{itiger: an automatic issue title generation tool}.
In: \bbtitle{Proceedings of the 30th ACM Joint European Software Engineering Conference and Symposium on the Foundations of Software Engineering},
pp. \bfpage{1637}--\blpage{1641}
(\byear{2022})
\end{bchapter}
\endbibitem

\bibitem[\protect\citeauthoryear{Liu et~al.}{2019}]{liu2019roberta}
\begin{botherref}
\oauthor{\bsnm{Liu}, \binits{Y.}},
\oauthor{\bsnm{Ott}, \binits{M.}},
\oauthor{\bsnm{Goyal}, \binits{N.}},
\oauthor{\bsnm{Du}, \binits{J.}},
\oauthor{\bsnm{Joshi}, \binits{M.}},
\oauthor{\bsnm{Chen}, \binits{D.}},
\oauthor{\bsnm{Levy}, \binits{O.}},
\oauthor{\bsnm{Lewis}, \binits{M.}},
\oauthor{\bsnm{Zettlemoyer}, \binits{L.}},
\oauthor{\bsnm{Stoyanov}, \binits{V.}}:
Roberta: A robustly optimized bert pretraining approach.
arXiv preprint arXiv:1907.11692
(2019)
\end{botherref}
\endbibitem

\bibitem[\protect\citeauthoryear{McCloskey and Cohen}{1989}]{mccloskey1989catastrophic}
\begin{bchapter}
\bauthor{\bsnm{McCloskey}, \binits{M.}},
\bauthor{\bsnm{Cohen}, \binits{N.J.}}:
\bctitle{Catastrophic interference in connectionist networks: The sequential learning problem}.
In: \bbtitle{Psychology of Learning and Motivation}
vol. \bseriesno{24},
pp. \bfpage{109}--\blpage{165}.
\bpublisher{Elsevier}, \blocation{???}
(\byear{1989})
\end{bchapter}
\endbibitem

\bibitem[\protect\citeauthoryear{Chen et~al.}{2020}]{chen2020stay}
\begin{bchapter}
\bauthor{\bsnm{Chen}, \binits{S.}},
\bauthor{\bsnm{Xie}, \binits{X.}},
\bauthor{\bsnm{Yin}, \binits{B.}},
\bauthor{\bsnm{Ji}, \binits{Y.}},
\bauthor{\bsnm{Chen}, \binits{L.}},
\bauthor{\bsnm{Xu}, \binits{B.}}:
\bctitle{Stay professional and efficient: automatically generate titles for your bug reports}.
In: \bbtitle{Proceedings of the 35th IEEE/ACM International Conference on Automated Software Engineering},
pp. \bfpage{385}--\blpage{397}
(\byear{2020})
\end{bchapter}
\endbibitem

\bibitem[\protect\citeauthoryear{Lewis et~al.}{2020}]{lewis2020bart}
\begin{bchapter}
\bauthor{\bsnm{Lewis}, \binits{M.}},
\bauthor{\bsnm{Liu}, \binits{Y.}},
\bauthor{\bsnm{Goyal}, \binits{N.}},
\bauthor{\bsnm{Ghazvininejad}, \binits{M.}},
\bauthor{\bsnm{Mohamed}, \binits{A.}},
\bauthor{\bsnm{Levy}, \binits{O.}},
\bauthor{\bsnm{Stoyanov}, \binits{V.}},
\bauthor{\bsnm{Zettlemoyer}, \binits{L.}}:
\bctitle{Bart: Denoising sequence-to-sequence pre-training for natural language generation, translation, and comprehension}.
In: \bbtitle{Proceedings of the 58th Annual Meeting of the Association for Computational Linguistics},
pp. \bfpage{7871}--\blpage{7880}
(\byear{2020})
\end{bchapter}
\endbibitem

\bibitem[\protect\citeauthoryear{Sutskever et~al.}{2014}]{sutskever2014sequence}
\begin{botherref}
\oauthor{\bsnm{Sutskever}, \binits{I.}},
\oauthor{\bsnm{Vinyals}, \binits{O.}},
\oauthor{\bsnm{Le}, \binits{Q.V.}}:
Sequence to sequence learning with neural networks.
Advances in neural information processing systems
\textbf{27}
(2014)
\end{botherref}
\endbibitem

\end{thebibliography}
\end{document}